\documentclass[prd,aps, tightenlines, showpacs,nofootinbib,superscriptaddress,notitlepage]{revtex4-1}
\usepackage{amssymb}
\usepackage{amsmath}
\usepackage{epsfig}

\begin{document}

\title{Inclusive Hadron Productions in $pA$ Collisions}
\author{Giovanni A. Chirilli}
\affiliation{Nuclear Science Division, Lawrence Berkeley National Laboratory, Berkeley,
CA 94720, USA}
\author{Bo-Wen Xiao}
\affiliation{Department of Physics, Pennsylvania State University, University Park, PA
16802, USA}
\author{Feng Yuan}
\affiliation{Nuclear Science Division, Lawrence Berkeley National Laboratory, Berkeley,
CA 94720, USA}

\begin{abstract}
We calculate inclusive hadrons production in $pA$ collisions in the small-$x$ saturation
formalism at one-loop order. The differential cross section is written
into a factorization form in the coordinate space at the next-to-leading order, while the naive
form of the convolution in the transverse momentum space does not
hold. The rapidity divergence with small-$x$ dipole gluon
distribution of the nucleus is factorized into the energy
evolution of the dipole gluon distribution function, which is
known as the Balitsky-Kovchegov equation. Furthermore, the
collinear divergences associated with the incoming parton
distribution of the nucleon and the outgoing fragmentation
function of the final state hadron are factorized into the
splittings of the associated parton distribution and fragmentation
functions, which allows us to reproduce the well-known DGLAP
equation. The hard coefficient function, which is finite and free
of divergence of any kind, is evaluated at one-loop order.
\end{abstract}

\maketitle

\section{Introduction}

Inclusive hadron production in $pA$ collisions have attracted much of
theoretical interests in recent years~\cite{Dumitru:2002qt,Kharzeev:2003wz,
Albacete:2003iq,Blaizot:2004wu,Baier:2005dz,Dumitru:2005gt,JalilianMarian:2005jf,
Albacete:2010bs,Altinoluk:2011qy,Qiu:2004da, Guzey:2004zp,Kopeliovich:2005ym,Frankfurt:2007rn}.
In particular, the suppression of hadron production in the forward $dAu$ scattering at RHIC observed in the
experiments~\cite{Arsene:2004ux,Adams:2006uz} has been regarded as one of the evidences
for the gluon saturation at small-$x$ in a large nucleus~\cite{JalilianMarian:2005jf,Albacete:2010bs,McLerran:2011zz}.
Saturation phenomenon at small-$x$ in nucleon and nucleus plays an important role
in high energy hadronic scattering~\cite{Gribov:1984tu,Mueller:1985wy,McLerran:1993ni,arXiv:1002.0333}.
In this paper, as an important step toward a complete description of hadron
production in $pA$ collisions in the saturation formalism, we calculate the one-loop perturbative
corrections. Previous attempts have been made in the literature.
In particular, in Ref.~\cite{Dumitru:2005gt}, part of one-loop
diagrams were evaluated. However, the rapidity divergence is not identified
and the collinear evolution effects are not complete. Recently, some of the higher
order corrections were discussed in Ref.~\cite{Altinoluk:2011qy}, where
it was referred as ``inelastic" contribution. In the following,
we will calculate the complete next-to-leading order (NLO) corrections to
this process in the saturation formalism. A brief summary of our results has been published
earlier in Ref.~\cite{Chirilli:2011km}.

Inclusive hadron production in $pA$ collisions,
\begin{equation}
p+A\to h+X \ ,\label{pA}
\end{equation}
can be viewed as a process where a parton from the nucleon
(with momentum $p$) scatters on the nucleus target
(with momentum $P_A$), and fragments into final state hadron with momentum $%
P_h$. In the dense medium of the large nucleus and at small-$x$,
the multiple interactions become important, and we need to perform
the relevant resummation to make the reliable predictions.
This is particularly important because the final
state parton is a colored object. Its interactions with
the nucleus target before it fragments into the hadron is crucial
to understand the nuclear effects in this process.
In our calculations, we follow the high energy factorization, also called
color-dipole or color-glass-condensate (CGC),
formalism~\cite{CU-TP-441a,Balitsky:1995ub,arXiv:1002.0333}
to evaluate the above process up to one-loop order.
We notice that alternative approaches have been proposed in the
literature~\cite{Qiu:2004da, Guzey:2004zp,Kopeliovich:2005ym,Frankfurt:2007rn}
to calculate the nuclear effects in this process.

According to our calculations, the QCD factorization formalism for the above
process reads as,
\begin{eqnarray}
\frac{d^3\sigma^{ p+A\to h+X}}{dyd^2p_\perp}&=&\sum\limits_a \int \frac{dz}{%
z^2}\frac{dx}{x} \xi xf_a(x,\mu) D_{h/c}(z,\mu) \int [dx_\perp]
S_{a,c}^Y([x_\perp])\mathcal{H}_{a\to
c}(\alpha_s,\xi,[x_\perp]\mu) \ , \label{fac}
\end{eqnarray}
where $\xi=\tau/xz$ with $\tau=p_\perp e^y/\sqrt{s}$, $y$ and
$p_\perp$ the rapidity and transverse momentum for the final state
hadron and $s$ the total center of mass energy square
$s=(p+P_A)^2$, respectively. Schematically, this factorization
is illustrated in Fig.~1, where the incoming parton
described by the parton distribution $f_a(x)$ scatters off the
nuclear target represented by multiple-point correlation function
$S^Y([x_\perp])$, and fragments into the final state hadron defined
by the fragmentation function $D_{h/c}(z)$. All these quantities
have clear operator definitions in QCD. In particular,
$f_a(x)$ and $D_{h/c}(z)$ are collinear parton distribution and fragmentation
functions which only depend on the longitudinal momentum fraction $x$ of
the nucleon carried by the parton $a$, and the momentum fraction $z$ of
parton $c$ carried by the final state hadron $h$, respectively.
From the nucleus side, it is the multi-point correlation functions
denoted as $S_{a,c}^Y(x_\perp)$ (see the definitions below) that enters
in the factorization formula, depending on the flavor of the
incoming and outgoing partons and the gluon rapidity $Y$ associated with
the nucleus: $Y\approx \ln(1/x_g)$ with $x_g$ being longitudinal momentum
fraction.

\begin{figure}[tbp]
\begin{center}
\includegraphics[width=7cm]{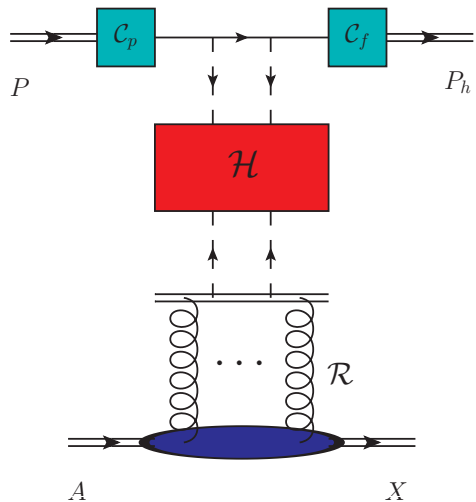}
\end{center}
\caption[*]{Schematic plot of the factorization, where $\mathcal{H}$
indicates the hard factor, $\mathcal{R}$ represents the rapidity
divergence which are factorized into the dipole gluon distribution
of the target nucleus ($A$), $\mathcal{C}_p$ and $\mathcal{C}_f$
stand for the collinear divergences which are absorbed into the
parton distribution functions of the projectile proton ($P$) and
hadron ($P_h$) fragmentation functions, respectively.}
\label{factorizationp}
\end{figure}

At the leading order, $S^Y$ represent the two-point functions, including
the dipole gluon distribution functions in the elementary and
adjoint representations for the quark and gluon initialed
subprocesses~\cite{arXiv:1002.0333}, respectively. Higher order
corrections will have terms that depend on the correlation
functions beyond the simple two-point functions. Because of this
reason, the integral $[dx_\perp]$ represents all possible
integrals at the particular order.

To evaluate NLO corrections, we will calculate the gluon radiation
contributions. At one-loop order, the gluon radiation will introduce
various divergences. The factorization formula in Eq.~(\ref{fac})
is to factorize these divergences into the relevant factors.
For example, there will be collinear divergences associated with the incoming
parton distribution and final state fragmentation functions.
In addition, there is the rapidity divergence associated with
$S^Y([x_\perp])$. These divergences naturally show up in
higher order calculations. The idea of the factorization is to
demonstrate that these divergences can be absorbed into the various
factors in the factorization formula. After subtracting these divergences,
we will obtain the hard factors
$\mathcal{H}_{a\to c}$, which describes the partonic scattering amplitude
of parton $a$ into a parton $c$ in the dense medium. This hard
factor includes all order perturbative corrections, and can be
calculated order by order. Although there is no simple
$k_\perp$-factorization form beyond leading order
formalism, we will find that in the
coordinate space, the cross section can be written into a nice
factorization form as Eq.~(\ref{fac}). Besides the explicit
dependence on the variables shown in Eq.~(\ref{fac}), there are
implicit dependences on $p_\perp [x_\perp]$ in the hard
coefficients as well.

Two important variables are introduced to separate different
factorizations for the physics
involved in this process: the collinear factorization scale $\mu$
and the energy evolution rapidity dependence $Y$. The physics
associated with $\mu$ follows the normal collinear QCD
factorization, whereas the rapidity factorization $Y$ takes into
account the small-$x$ factorization. The evolution respect to
$\mu$ is controlled by the usual DGLAP evolution, whereas that for
$S_a^Y$ by the Balitsky-Kovchegov (BK)
evolution~\cite{Balitsky:1995ub,Kovchegov:1999yj}. In general, the energy evolution of any correlation functions can be given by
the JIMWLK equation\cite{Jalilian-Marian:1997jx+X}, and the resulting equation is equivalent to the BK equation for dipole amplitudes. In particular,
our one-loop calculations will demonstrate the important
contribution from this rapidity divergence.
Schematically, this factorization is shown in
Fig.~\ref{factorizationp}.

Our calculations should be compared to the similar calculations
at next-to-leading order for the DIS structure functions in the saturation
formalism~\cite{Balitsky:2010ze,Beuf:2011xd,Al}. All these calculations
are important steps to demonstrate the factorization for general hard
processes in the small-$x$ saturation formalism~\cite{Gelis:2008rw}.
The rest of the paper is organized as follows. In Sec. II, we
discuss the leading order results for inclusive hadron production
in $pA$ collision, where we also set up the framework for the NLO
calculations. Sec. III. is divided into four
subsections, in which we calculate the NLO cross section for the
$q\to q$, $g\to g$, $q\to g$ and $g\to q$ channels . The summary and further discussions are given in Sec.
IV.

\section{The leading order single inclusive cross section.}

\begin{figure}[tbp]
\begin{center}
\includegraphics[width=8cm]{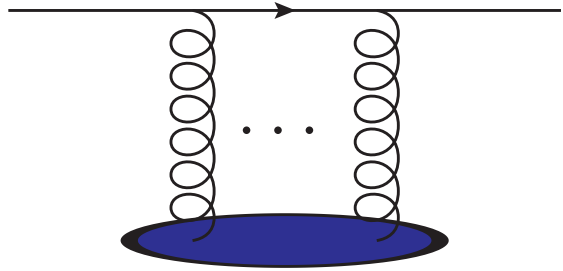}
\end{center}
\caption[*]{Typical Feynman diagrams for the leading order quark production $%
qA\rightarrow q+X$.}
\label{lof}
\end{figure}

The leading order result was first formulated in Ref.~\cite{Dumitru:2002qt}.
For the purpose of completeness,  we briefly derive the leading
order cross section to set up the baseline for the NLO calculation.
Let us begin with the quark channel in $pA$ collisions.
As illustrated in Fig.~\ref{lof}, the multiple scattering between the quark
from the proton and the dense gluons inside the nucleus target
can be cast into the Wilson line
\begin{equation}
U(x_\perp)=\mathcal{P}\exp\left\{ig_S\int_{-\infty}^{+\infty} \text{d}%
x^+\,T^cA_c^-(x^+,x_\perp)\right\} \ ,
\end{equation}
with $A_c^-(x^+,x_\perp)$ being the gluon field solution of the classical
Yang-Mills equation inside the large nucleus target.

Therefore, the leading-order cross section for producing a quark with finite
transverse momentum $k_{\perp }$ at rapidity $y$ in the channel $%
qA\rightarrow qX$ can be written as:
\begin{equation}
\frac{d\sigma _{\text{LO}}^{pA\rightarrow qX}}{d^{2}k_{\perp }dy}%
=\sum_{f}xq_{f}(x)\int \frac{d^{2}x_{\perp }d^{2}y_{\perp }}{(2\pi )^{2}}%
e^{-ik_{\perp }\cdot (x_{\perp }-y_{\perp })}\frac{1}{N_{c}}\left\langle
\text{Tr}U(x_{\perp })U^{\dagger }(y_{\perp })\right\rangle _{Y},
\end{equation}%
with $x=\frac{k_{\perp }}{\sqrt{s}}e^{y}$ and $x_{g}=\frac{k_{\perp }}{\sqrt{%
s}}e^{-y}$. The notation $\langle \dots \rangle _{Y}$ indicates the CGC
average of the color charges over the nuclear wave function where $Y\simeq
\ln 1/x_{g}$ and $x_{g}$ is the smallest longitudinal momentum fraction of
the probed gluons, and is determined by the kinematics~\footnote{%
Here we are only interested in the inelastic production of the quark in the
forward scattering which produces quark with finite transverse momentum.
There is also elastic scattering contribution to the cross section which
generates vanishing $k_{\perp }$, such as $\sum_{f}xq_{f}(x)%
\delta ^{(2)}(k_{\perp })\int d^{2}b$ to the total cross section.}. Normally, we first compute the correlator $\left\langle
\text{Tr}U(x_{\perp })U^{\dagger }(y_{\perp })\right\rangle$ in the McLerran-Venugopalan model\cite{McLerran:1993ni} as the initial condition, and then
we perform the energy evolution for the correlator which introduces the rapidity ($Y$) dependence. The energy evolution equation at small-$x$ for dense nucleus targets
is the BK equation as we shall demonstrate later when we remove the rapidity divergence.
When multiplied by the fragmentation function, the above result will
lead to the differential cross section for hadron production in $pA$ collisions.

It is straightforward to include the gluon initiated channel, and the full
leading order hadron production cross section can be written as
\begin{equation}
\frac{d\sigma _{\text{LO}}^{pA\rightarrow hX}}{d^{2}p_{\perp }dy_{h}}%
=\int_{\tau}^{1}\frac{dz}{z^{2}}\left[\sum_{f}x_{p}q_{f}(x_{p})\mathcal{F}%
(k_\perp)D_{h/q}(z)+x_{p}g(x_{p})\tilde{\mathcal{F}}(k_\perp)D_{h/g}(z)%
\right],  \label{lo}
\end{equation}%
with $p_{\perp }=zk_{\perp }$, $x_{p}=\frac{p_{\perp }}{z\sqrt{s}}e^{y_{h}}$%
, $\tau=zx_{p}$ and $x_{g}=\frac{p_{\perp }}{z\sqrt{s}}e^{-y_{h}}$. Here we
have defined
\begin{equation}
\mathcal{F}(k_\perp)=\int\frac{d^2x_\perp d^2y_\perp}{(2\pi)^2}%
e^{-ik_\perp\cdot (x_\perp-y_\perp)}S_Y^{(2)}(x_\perp,y_\perp),  \label{fd}
\end{equation}
with $S_Y^{(2)}(x_\perp,y_\perp)=\frac{1}{N_{c}}\left\langle \text{Tr}%
U(x_{\perp})U^{\dagger }(y_{\perp })\right\rangle _{Y}$. $\tilde{\mathcal{F}}%
(k_\perp)$ is defined similarly but in the adjoint representation
\begin{equation}
\tilde{\mathcal{F}}(k_\perp)=\int\frac{d^2x_\perp d^2y_\perp}{(2\pi)^2}%
e^{-ik_\perp\cdot
(x_\perp-y_\perp)}\tilde{S}_Y^{(2)}(x_\perp,y_\perp), \label{fda}
\end{equation}
where $\tilde{S}_Y^{(2)}(x_\perp,y_\perp)=\frac{1}{N_{c}^2-1}\left\langle \text{Tr}%
W(x_{\perp})W^{\dagger }(y_{\perp })\right\rangle _{Y}$ and $W(x)$
is a Wilson line in the adjoint representation. It represents
the multiple interaction between the final state gluon and the
nucleus target. In general, the adjoint Wilson lines can be
replaced by two fundamental Wilson lines by using the identity
\begin{equation}
W^{ab}(x_\perp)=2\text{Tr}\left[T^aU(x_\perp)T^bU^\dagger(x_\perp)\right],
\label{adjoint}
\end{equation}
and the color matrices can be removed using the Fierz identity $%
T^a_{ij}T^a_{kl}=\frac{1}{2}\delta_{il}\delta_{jk}-\frac{1}{2N_c}%
\delta_{ij}\delta_{kl}$. It is straightforward to show that
\begin{equation}
\tilde{S}_Y^{(2)}(x_\perp,y_\perp)=\frac{1}{N_{c}^2-1}\left[\left\langle \text{Tr}%
U(x_{\perp})U^{\dagger }(y_{\perp
})\textrm{Tr}U(y_{\perp})U^{\dagger }(x_{\perp })\right\rangle
_{Y}-1\right],
\end{equation}
which, in the large $N_c$ limit, allows us to write
\begin{equation}
\tilde{\mathcal{F}}(k_\perp)=\int\frac{d^2x_\perp d^2y_\perp}{(2\pi)^2}%
e^{-ik_\perp\cdot
(x_\perp-y_\perp)}S_Y^{(2)}(x_\perp,y_\perp)S_Y^{(2)}(y_\perp,x_\perp)\ .
\label{fda2}
\end{equation}
It is very important to keep in mind that the normalization of the dipole
amplitudes $S^{(2)}(x_\perp,y_\perp)$ is unity when $x_\perp=y_\perp$.
In addition, since normally
$\left\langle \text{Tr}U(x_{\perp})U^{\dagger }(y_{\perp })\right\rangle _{Y}$
is real, it is easy to see that $S^{(2)}(x_\perp,y_\perp)=S^{(2)}(y_\perp,x_\perp)$.
If we further neglect the impact parameter dependence, one will find that
$S^{(2)}(x_\perp,y_\perp)=\exp \left[-\frac{Q_s^2 (x_\perp-y_\perp)^2}{4}\right]$
in the McLerran-Venugopalan model, where $Q_s$ is the saturation momentum which characterizes the
density of the target nucleus. The analytical form of the dipole amplitude
can help us to test the properties of dipole amplitudes mentioned above.

We would like to emphasize that in Eq.~(\ref{lo}) we do not include the
transverse momentum dependence in the incoming parton distribution
from the nucleon. In the forward $pA$ collisions, the transverse momentum
dependence from the incoming parton distribution of the nucleon is not
as important as that from the nucleus target. Therefore, in the current
calculations, we neglect this effect. As a consistent check, the one-loop
calculations in the following support this assumption. In particular, the
collinear divergence associated with the incoming parton distribution
contains no transverse momentum dependence.

\section{The Next-to-leading order cross section}

In this section, we will present the detailed calculations for the NLO
corrections to the leading order result in Eq.~(\ref{lo}). There are
four partonic channels: $q\to qg$, $g\to gg$, $q\to gq$, $g\to q\bar q$.
We will carry out the calculations for these channels separately.

\subsection{The quark channel $q\to q$}

\begin{figure}[tbp]
\begin{center}
\includegraphics[width=10cm]{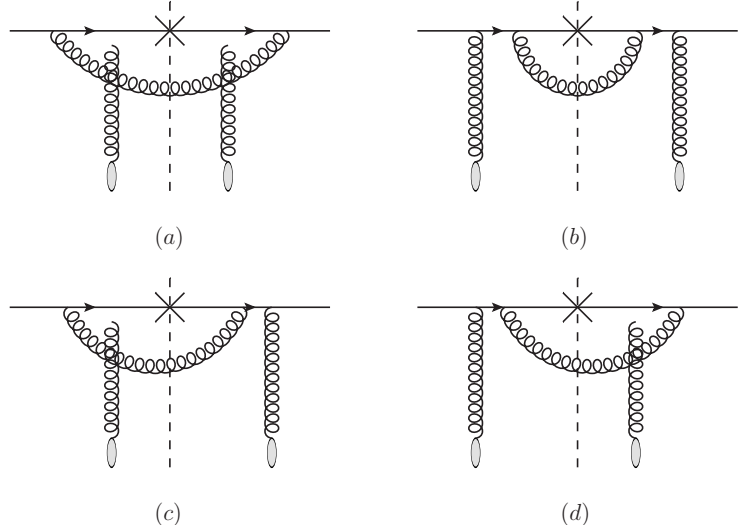}
\end{center}
\caption[*]{The real diagrams for the next-to-leading order quark production
$qA\rightarrow q+X$.}
\label{nloqr}
\end{figure}

The quark production contribution contains the real and virtual gluon
radiation at the NLO. For the real contribution, we will calculate $q\rightarrow qg$ first.
The real diagrams with a quark (with transverse coordinate $b_\perp$)
and gluon (with transverse coordinate $x_\perp$) in the final state,
as shown in Fig.~\ref{nloqr}, have been studied in
Ref.~~\cite{hep-ph/0405266, Dominguez:2010xd, Dominguez:2011wm}. We take eq.(78) of Ref.~\cite%
{Dominguez:2011wm} as our starting point which gives\footnote{%
For convention reasons, we have interchanged the definition of $z$ and $1-z$
and replaced the variable $z$ by $\xi$.}
\begin{eqnarray}
\frac{d\sigma ^{qA\rightarrow qg X}}{d^3k_1d^3k_2}&=&\alpha
_{S}C_F\delta(p^+-k_1^+-k_2^+) \int \frac{\text{d}^{2}x_{\perp}}{(2\pi)^{2}}%
\frac{\text{d}^{2}x_{\perp}^{\prime }}{(2\pi )^{2}}\frac{\text{d}%
^{2}b_{\perp}}{(2\pi)^{2}}\frac{\text{d}^{2}b_{\perp}^{\prime }}{(2\pi )^{2}}
\notag \\
&&\times e^{-ik_{1\perp }\cdot(x_{\perp}-x^{\prime }_{\perp})}e^{-ik_{2\perp
}\cdot (b_{\perp}-b_{\perp}^{\prime })} \sum_{\lambda\alpha\beta}
\psi^{\lambda\ast}_{\alpha\beta}(u^{\prime}_{\perp})\psi^\lambda_{\alpha%
\beta}(u_{\perp})  \notag \\
&&\times \left[S^{(6)}_{Y}(b_{\perp},x_{\perp},b^{\prime}_{\perp},x^{%
\prime}_{\perp})+S^{(2)}_{Y}(v_{\perp},v^{\prime}_{\perp})\right.  \notag \\
&&\quad\left.-S^{(3)}_{Y}(b_{\perp},x_{\perp},v^{\prime}_{%
\perp})-S^{(3)}_{Y}(v_{\perp},x^{\prime}_{\perp},b^{\prime}_{\perp})\right].
\label{partqqg}
\end{eqnarray}
where $u_{\perp}=x_{\perp}-b_{\perp}$, $u^{\prime}_{\perp}=x^{\prime}_{%
\perp}-b^{\prime}_{\perp}$, $v_{\perp}=(1-\xi)x_{\perp}+\xi b_{\perp}$, $%
v^{\prime}_{\perp}=(1-\xi)x^{\prime}_{\perp}+\xi b^{\prime}_{\perp}$ and
\begin{align}
S^{(6)}_{Y}(b_{\perp},x_{\perp},b^{\prime}_{\perp},x^{\prime}_{\perp})&=%
\frac{1}{C_FN_c}\left\langle\text{Tr}\left(U(b_{\perp})U^\dagger(b^{%
\prime}_{\perp})T^dT^c\right)\left[W(x_{\perp})W^\dagger(x^{\prime}_{\perp})%
\right]^{cd}\right\rangle_{Y}, \\
S^{(3)}_{Y}(b_{\perp},x_{\perp},v^{\prime}_{\perp})&=\frac{1}{C_FN_c}%
\left\langle\text{Tr}\left(U(b_{\perp})T^dU^\dagger(v^{\prime}_{\perp})T^c%
\right)W^{cd}(x_{\perp})\right\rangle_{Y}.
\end{align}
For a right-moving massless quark, with initial longitudinal momentum $p^+$
and no transverse momentum, the splitting wave function in transverse
coordinate space is given by
\begin{equation}
\psi^\lambda_{\alpha\beta}(p^+,k_1^+,r_{\perp})=2\pi i\sqrt{\frac{2}{k_1^+}}%
\begin{cases}
\frac{r_{\perp}\cdot\epsilon^{(1)}_\perp}{r_{\perp}^2}(\delta_{\alpha-}%
\delta_{\beta-}+\xi\delta_{\alpha+}\delta_{\beta+}), & \lambda=1, \\
\frac{r_{\perp}\cdot\epsilon^{(2)}_\perp}{r_{\perp}^2}(\delta_{\alpha+}%
\delta_{\beta+}+\xi\delta_{\alpha-}\delta_{\beta-}), & \lambda=2.%
\end{cases}
\ ,  \label{wvfunction}
\end{equation}
where $\lambda$ is the gluon polarization, $\alpha,\beta$ are
helicities for the incoming and outgoing quarks, and $1-\xi=\frac{k_1^+}{p^+}
$ is the momentum fraction of the incoming quark carried by the gluon. Since
the Wilson lines in the fundamental representation and the adjoint
representation resum the multiple interactions of quarks and gluons with the
nucleus target, respectively, one can easily see that these four terms in
the last two lines of the Eq.~(\ref{partqqg}) correspond to those four
graphs in Fig.~\ref{nloqr}. The $S^{(6)}_{Y}$ term which corresponds to Fig.~%
\ref{nloqr} (a) and resums all the multiple interactions between the
quark-gluon pair and the nucleus target, represents the case where
interactions take place after the splitting both in the amplitude and in the
conjugate amplitude. The $S^{(2)}_{Y}$ term which comes from Fig.~\ref{nloqr}
(b), resums the interactions before the splitting only and the $S^{(3)}_{Y}$
terms represent the interference terms as shown in Fig.~\ref{nloqr} (c) and
(d).

There are two contributions for inclusive hadron production at the
next-to-leading order, namely, quark productions associated with $D_{h/q}$
which is indicated by the cross in Fig.~\ref{nloqr} (while the gluon is
integrated) and gluon productions associated with the fragmentation function
$D_{h/g}$ (while the quark is integrated).

Let us study the former case by integrating over the phase space of
the final state gluon $(k^+_1, k_{1\perp})$. We can cast the real contribution into
\begin{eqnarray}
&&\frac{\alpha _{s}}{2\pi ^{2}}\int \frac{dz}{z^{2}}D_{h/q}(z)\int_{\tau
/z}^{1}d\xi \frac{1+\xi ^{2}}{1-\xi }xq(x)\left\{ C_{F}\int d^{2}k_{g\perp }%
\mathcal{I}(k_\perp,k_{g\perp})\right.  \notag \\
&&\left. ~~~+\frac{N_{c}}{2}\int d^{2}k_{g\perp }d^{2}{k_{g1\perp }}\mathcal{%
J}(k_\perp,k_{g\perp},k_{g1\perp})\right\} \ ,  \label{real}
\end{eqnarray}%
where $x=\tau /z\xi $ and $C_F=(N_c^2-1)/2N_c$, and $\mathcal{I}$ and $%
\mathcal{J}$ are defined as
\begin{eqnarray}
\mathcal{I}(k_\perp,k_{g\perp})&=&\mathcal{F}(k_{g\perp})\left[ \frac{%
k_\perp-k_{g\perp}}{( k_{\perp }-k_{g\perp})^{2}}-\frac{k_\perp-\xi
k_{g\perp}}{(k_{\perp }-\xi k_{g\perp})^{2}}\right]^2,  \notag \\
\mathcal{J}(k_\perp,k_{g\perp},k_{g1\perp})&=&\left[ \mathcal{F}%
(k_{g\perp})\delta ^{\left( 2\right) }\left( k_{g1\perp }-k_{g\perp}\right) -%
\mathcal{G}(k_{g\perp},k_{g1\perp })\right]\frac{2(k_{\perp }-\xi
k_{g\perp})\cdot (k_{\perp }-k_{g1\perp })}{(k_{\perp }-\xi
k_{g\perp})^{2}(k_{\perp }-k_{g1\perp })^{2}} \ ,  \notag \\
\text{with} \quad \mathcal{G}(k_\perp,l_{\perp})&=&\int\frac{d^2x_\perp
d^2y_\perp d^2b_\perp}{(2\pi)^4}e^{-ik_\perp\cdot
(x_\perp-b_\perp)-il_{\perp}\cdot(b_\perp-y_\perp)}
S_Y^{(4)}(x_\perp,b_\perp,y_\perp),
\end{eqnarray}
and $S_Y^{(4)}(x_\perp,b_\perp,y_\perp)=\frac{1}{N_c^2}\langle \mathrm{Tr}[{U}%
(x_\perp){U}^\dagger(b_\perp)] \mathrm{Tr}[{U}(b_\perp){U}%
^\dagger(y_\perp)]\rangle_Y$. Several steps are necessary in deriving the
above result from Eq.~(\ref{partqqg}). By integrating over the gluon
momentum, we identify $x_\perp$ to $x^{\prime}_{\perp}$ which simplifies $%
S^{(6)}_{Y}$ to $S^{(2)}(b_\perp, b^{\prime}_\perp)$. This is expected since
we know the multiple interactions between the gluon and the nucleus target
should cancel if the gluon is not observed. Furthermore, using the Fierz
identity, one can write
\begin{equation}
S^{(3)}_{Y}(b_\perp,x_\perp,v^{\prime}_\perp)=\frac{N_c}{2C_F}\left[%
S_Y^{(4)}(b_\perp,x_\perp,v^{\prime}_\perp)-\frac{1}{N_c^2}S_Y^{(2)}(b_\perp,v^{%
\prime}_\perp)\right],
\end{equation}
which only involves the Wilson lines in the fundamental representation.
Then, the final steps, which include the Fourier transforms, as well as the
convolutions of the quark distribution and fragmentation function, are quite
straightforward.

Before we proceed to the calculations of the virtual diagrams, we comment on
the result shown in Eq.~(\ref{real}). The major obstacles of evaluating the
integrals in Eq.~(\ref{real}) are the divergences. There are three types of
singularities lying in that equation, namely, the rapidity divergence which
occurs at $\xi =1$ when the rapidity of the radiated gluon becomes $-\infty $, and
the collinear singularities which correspond to the cases that the final
state gluon is either collinear to the initial quark or final state quark.
We shall expect that the virtual diagrams cancel some part of the
divergences, while the uncancelled divergences shall be absorbed into the
renormalization of the quark distribution and fragmentation functions as
well as the target dipole gluon distribution ($S_{Y}^{(2)}(x_{\perp
},y_{\perp })$). After these subtractions, the remainder contributions
should be finite and give us the NLO correction to the single inclusive
hadron production cross section.

\begin{figure}[tbp]
\begin{center}
\includegraphics[width=10cm]{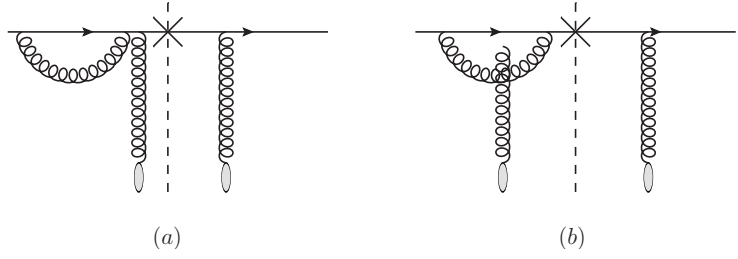}
\end{center}
\caption[*]{Typical virtual diagrams for the next-to-leading order quark
production $qA\rightarrow q+X$.}
\label{nloqv}
\end{figure}

The evaluation of the virtual graphs as shown in Fig.~\ref{nloqv} are quite
simple in the dipole picture. Their contributions are proportional to
\begin{eqnarray}
&&-2\alpha _{s}C_{F}\int \frac{d^{2}v_\perp}{(2\pi )^{2}}\frac{%
d^{2}v^{\prime }_\perp}{(2\pi )^{2}}\frac{d^{2}u_\perp}{(2\pi )^{2}}%
e^{-ik_{\perp }\cdot (v_\perp-v^{\prime
}_{\perp})}\sum_{\lambda\alpha\beta}\psi^{\lambda\ast}_{\alpha\beta}(u_%
\perp)\psi^{\lambda}_{\alpha\beta} (u_\perp)  \notag \\
&&\times \left[ S_Y^{(2)}(v_\perp,v_\perp^{\prime
})-S^{(3)}_{Y}(b_{\perp},x_{\perp},v^{\prime}_{\perp})\right] ,  \label{v1}
\end{eqnarray}
where the factor of $2$ takes care of the fact that the mirror diagrams of
Fig.~\ref{nloqv} give the identical contributions when the virtual loop is
on the right side of the cut. It is straightforward to see that these two
terms in the last line of Eq.~(\ref{v1}) correspond to the Fig.~\ref{nloqv}
(a) and (b), respectively. This eventually leads to
\begin{eqnarray}
&&-\frac{\alpha _{s}}{2\pi ^{2}}\int \frac{dz}{z^{2}}D_{h/q}(z)
x_pq(x_p)\int_{0}^{1}d\xi \frac{1+\xi ^{2}}{1-\xi }  \notag \\
&&\times \left\{ C_{F}\int d^{2}q_{\perp }\mathcal{I} (q_\perp,k_\perp) +%
\frac{N_{c}}{2}\int d^{2}q_{\perp }d^{2}{k_{g1\perp }}\mathcal{J}%
(q_\perp,k_\perp,k_{g1\perp}) \right\}\ ,  \label{virtual}
\end{eqnarray}
where explicitly one writes
\begin{eqnarray}
\mathcal{I}(q_\perp,k_{\perp})&=&\mathcal{F}(k_{\perp})\left[ \frac{%
q_\perp-k_{\perp}}{( q_{\perp }-k_{\perp})^{2}}-\frac{q_\perp-\xi k_{\perp}}{%
(q_{\perp }-\xi k_{\perp})^{2}}\right]^2,  \notag \\
\mathcal{J}(q_\perp,k_\perp,k_{g1\perp})&=&\left[ \mathcal{F}(k_{\perp
})\delta ^{\left( 2\right) }\left( k_{g1\perp }-k_{\perp }\right) -\mathcal{G%
}(k_{\perp },k_{g1\perp })\right]\frac{2(q_{\perp }-\xi k_{\perp })\cdot
(q_{\perp }-k_{g1\perp })}{(q_{\perp }-\xi k_{\perp })^{2}(q_{\perp
}-k_{g1\perp })^{2}} .
\end{eqnarray}
It is easy to see that the virtual contributions indeed contain three types
of singularities as we mentioned before. There are two important features
that we wish to emphasize here. First, the rapidity divergence term is only
proportional to $N_c/2$ since $\mathcal{I}$ vanishes at $\xi\to 1$ limit.
This agrees with the BK equation since there is no $1/N_c^2$ corrections to
the leading order BK equation. Second, when one integrates over the quark
transverse momentum $k_\perp$, the rapidity divergence disappears due to the
complete cancellation between the real and virtual contributions.

\subsubsection{The rapidity divergence}

Now we are ready to evaluate NLO contributions by the following
procedures. First, we remove the rapidity divergence terms from the real and
virtual contributions by doing the following subtractions
\begin{eqnarray}
\mathcal{F}(k_{\perp }) &=&\mathcal{F}^{(0)}(k_{\perp })-\frac{\alpha
_{s}N_{c}}{2\pi ^{2}}\int_{0}^{1}\frac{d\xi }{1-\xi }\int \frac{%
d^{2}x_{\perp }d^{2}y_{\perp }d^{2}b_{\perp }}{(2\pi )^{2}}e^{-ik_{\perp
}\cdot (x_{\perp }-y_{\perp })}  \notag \\
&&\times \frac{(x_{\perp }-y_{\perp })^{2}}{(x_{\perp }-b_{\perp
})^{2}(y_{\perp }-b_{\perp })^{2}}\left[ S_{Y}^{(2)}(x_{\perp },y_{\perp
})-S_{Y}^{(4)}(x_{\perp },b_{\perp },y_{\perp })\right] \ ,  \label{BKg}
\end{eqnarray}%
where $\mathcal{F}^{(0)}(k_{\perp })$ is the bare dipole gluon distribution
which appears in the leading order cross section as in Eq.~(\ref{lo}) and it
is divergent. $\mathcal{F}(k_{\perp })$ is the renormalized dipole gluon
distribution and it is assumed to be finite. We can always decompose the
dipole splitting kernel as
\begin{equation}
\frac{(x_{\perp }-y_{\perp })^{2}}{(x_{\perp }-b_{\perp })^{2}(y_{\perp
}-b_{\perp })^{2}}=\frac{1}{(x_{\perp }-b_{\perp })^{2}}+\frac{1}{(y_{\perp
}-b_{\perp })^{2}}-\frac{2(x_{\perp }-b_{\perp })\cdot (y_{\perp }-b_{\perp
})}{(x_{\perp }-b_{\perp })^{2}(y_{\perp }-b_{\perp })^{2}},
\end{equation}%
where the first two terms are removed from the virtual
contribution while the last term is removed from the real
diagrams. This procedure is similar to that for the collinear
factorization, where we modify the bare leading order parton
distributions to the finite parton distribution with the higher
order radiation. Using Eqs.~(\ref{fd}, \ref{BKg}), we can
see that the differential change of the dipole amplitude $%
S_{Y}^{(2)}(x_{\perp },y_{\perp })$ yields the BK equation
\begin{equation}
\frac{\partial}{\partial Y}S_{Y}^{(2)}(x_{\perp },y_{\perp })=-\frac{\alpha
_{s}N_{c}}{2\pi ^{2}}\int \frac{d^{2}b_{\perp }\,(x_{\perp }-y_{\perp })^{2}%
}{(x_{\perp }-b_{\perp })^{2}(y_{\perp }-b_{\perp })^{2}}\left[
S_{Y}^{(2)}(x_{\perp },y_{\perp })-S_{Y}^{(4)}(x_{\perp },b_{\perp
},y_{\perp })\right] \ .  \label{BK}
\end{equation}%
It is important to note that if we conduct the leading order classical calculation, we will not get any energy dependence, namely the $Y$ dependence, in the scattering amplitudes.
It is the BK evolution equation as shown above which gives the energy dependence to those scattering amplitudes. To derive the BK equation from Eqs.~(\ref{fd}, \ref{BKg}), one needs to reset the upper
limit of the $d\xi $ integral in Eq.~(\ref{BKg}) to $ 1-e^{-Y}$, with $Y$ being the total rapidity gap between the projectile proton and the target nucleus. Here $Y \to \infty$ as the center of mass energy $s\to \infty$. By doing so, we introduce the rapidity $Y$ dependence, namely the
energy dependence, of the two-point function $S_{Y}^{(2)}(x_{\perp
},y_{\perp })$ from which the BK equation can be understood and
therefore derived. Another way to derive this equation is to
slightly move away from the light cone as in the derivation of the
Balitsky equation\cite{Balitsky:1995ub}. The rapidity divergence is an artifact that we put both the projectile and targets on the light cone in the high energy limit. By slightly tilting away from the light cone, we can modify the $\xi$ integral and obtain $\int_{0}^{1}\frac{d\xi }{1-\xi +e^{-Y}}$. In addition, when one integrates over
the transverse momentum $k_\perp$ as in Eq.~(\ref{BKg}), one finds
that the rapidity divergence disappears as
expected~\cite{Collins:2011zzd}.

The physical interpretation of the rapidity divergence subtraction is quite interesting. Although the soft gluon is emitted from the projectile proton which is moving on the forward light cone with the rapidity close to $+\infty$, it is easy to see that the rapidity of this soft gluon goes to $-\infty$ when $\xi\to 1$. As a matter of fact, this soft gluon can be regarded as collinear to the target nucleus which is moving on the backward light cone with the rapidity close to $-\infty$. Therefore, it is quite natural to renormalize this soft gluon into the gluon distribution function of the target nucleus through the BK evolution equation.

After the subtraction of the rapidity divergence, both of the real
and virtual contributions become regulated in terms of the $d\xi $
integral which leads to the change of the splitting function into
$\frac{1+\xi^2}{(1-\xi)_+}$. Here we introduce the following
property of the plus-function
\begin{equation}
\int_a^1d\xi \left(f(\xi)\right)_+ g(\xi)=\int_a^1d\xi f(\xi)[g(\xi)-g(1)]-g(1)\int^a_0
d\xi f(\xi),  \label{plus}
\end{equation}
where $g(\xi)$ can be any non-singular functions, while $f(\xi)$ is singular
at $\xi=1$ and $\left(f(\xi)\right)_+$ is regulated.

\subsubsection{The collinear divergence}

The second step is to use the dimensional regularization ($D=4-2\epsilon$)
and follow the $\overline{\mathrm{MS}}$ subtraction scheme, in order to
compute and remove the collinear divergence from both real and virtual
contributions. For convenience, we introduce the following integrals,
\begin{eqnarray}
I_{1}(k_{\perp }) &=&\int \frac{d^{2}k_{g\perp }}{(2\pi )^{2}}\mathcal{F}%
(k_{g\perp })\frac{1}{(k_{\perp }-k_{g\perp })^{2}}\ ,  \notag \\
I_{2}(k_{\perp }) &=&\int \frac{d^{2}k_{g\perp }}{(2\pi )^{2}}\mathcal{F}%
(k_{g\perp })\frac{(k_{\perp }-k_{g\perp })\cdot (k_{\perp }-\xi k_{g\perp })%
}{(k_{\perp }-k_{g\perp })^{2}(k_{\perp }-\xi k_{g\perp })^{2}}\ ,  \notag \\
I_{3}(k_{\perp }) &=&\int \frac{d^{2}k_{g\perp }d^{2}k_{g1\perp }}{(2\pi
)^{2}}\mathcal{G}(k_{g\perp },k_{g1\perp })\frac{(k_{\perp }-k_{g1\perp
})\cdot (k_{\perp }-\xi k_{g\perp })}{(k_{\perp }-k_{g1\perp })^{2}(k_{\perp
}-\xi k_{g\perp })^{2}}\ .
\end{eqnarray}%
Clearly, there is no divergence in $I_{3}$. Let us take the
evaluation of $I_{1}(k_{\perp })$ as an example. As standard
procedure in the dimensional regularization ($D=4-2\epsilon$) and the $%
\overline{\mathrm{MS}}$ subtraction scheme, we change the integral $\int
\frac{d^{2}k_{g\perp }}{(2\pi )^{2}}$ into $\mu^{2\epsilon}\int \frac{%
d^{2-2\epsilon}k_{g\perp }}{(2\pi )^{2-2\epsilon}}$ where $\mu$ is the scale
dependence coming from the strong coupling $g$. Using Eq.~(\ref{fd}) and the
identity
\begin{equation}
\int \text{d}^{2-2\epsilon}q_\perp e^{-iq_\perp \cdot r_\perp} \frac{%
\mu^{2\epsilon}}{q_\perp^2}= \pi \left(\frac{\mu^2 r_\perp^2}{4\pi}%
\right)^\epsilon \Gamma(-\epsilon),
\end{equation}
together with the convention $\frac{1}{\hat{\epsilon}}=\frac{1}{\epsilon}%
-\gamma_E+\ln 4\pi$, we can find
\begin{equation}
I_{1}(k_{\perp }) =\frac{1}{4\pi}\int\frac{d^2x_\perp d^2y_\perp}{(2\pi)^2}%
e^{-ik_\perp\cdot r_\perp}S_Y^{(2)}(x_\perp,y_\perp)\left(-\frac{1}{\hat{%
\epsilon}}+\ln \frac{c_0^2}{\mu^2r_\perp^2}\right),
\end{equation}
where $c_0=2e^{-\gamma_E}$, $\gamma_E$ is the Euler constant and $r_\perp =
x_\perp -y_\perp$.

To evaluate $I_{2}(k_{\perp })$, we first rewrite it as
\begin{equation}
I_{2}(k_{\perp })=-\frac{1}{4\pi }\mathcal{F}(k_{\perp })\ln
(1-\xi)^2+I_{21}(k_\perp) \ ,
\end{equation}%
where $I_{21}$ is finite and defined as
\begin{eqnarray}
I_{21}(k_\perp)&=&\int \frac{d^{2}k_ {g\perp }}{(2\pi )^{2}}\left[ \mathcal{F%
}(k_{g\perp })\frac{(k_{\perp }-k_{g\perp })\cdot (k_{\perp }-\xi k_{g\perp
})}{(k_{\perp }-k_{g\perp })^{2}(k_{\perp }-\xi k_{g\perp })^{2}} \right.
\notag \\
&&-\left. \mathcal{F}(k_\perp)\frac{(k_{\perp }-k_{g\perp })\cdot (\xi
k_{\perp }-k_{g\perp })}{(k_{\perp }-k_{g\perp })^{2}(\xi k_{\perp
}-k_{g\perp })^{2}} -\mathcal{F}(k_\perp) \frac{k_{g\perp }\cdot (k_{\perp
}- k_{g\perp })}{k_{g\perp }^{2}(k_{\perp }- k_{g\perp })^{2}}\right] \ .
\end{eqnarray}%
The basic idea is to subtract a term which is proportional to $\ln(1-\xi)^2$
from $I_2$. It is quite straightforward to show that the last two terms in
the above equation give $\mathcal{F}(k_\perp)\ln(1-\xi)^2$ by using the
integral identity
\begin{equation}
\int \frac{d^{2}k_{g\perp }}{(2\pi )^{2}}\left[\frac{(k_{\perp }-k_{g\perp
})\cdot (\xi k_{\perp }-k_{g\perp })}{(k_{\perp }-k_{g\perp })^{2}(\xi
k_{\perp }-k_{g\perp })^{2}}-\frac{k_{g\perp }\cdot (k_{g\perp }- k_{\perp })%
}{k_{g\perp }^{2}(k_{g\perp }- k_{\perp })^{2}}\right]=\frac{1}{4\pi} \ln%
\frac{1}{(1-\xi)^2}.  \label{intlog}
\end{equation}

To compute and remove the collinear divergence in the virtual diagrams, one
needs to use the following integral
\begin{equation}
\mu ^{2\epsilon }\int \frac{d^{2-2\epsilon }l_{\perp }}{(2\pi )^{2-2\epsilon
}}\frac{\Delta ^{2}}{(l_{\perp }-\Delta )^{2}l_{\perp }^{2}}=\frac{1}{2\pi }%
\left( -\frac{1}{\hat{\epsilon}}+\ln \frac{\Delta ^{2}}{\mu ^{2}}\right) ,
\label{loop}
\end{equation}%
where the usual Feynman integral trick is used in the derivation. Therefore,
setting the quark distribution, the fragmentation function and the splitting
function aside, the virtual contribution can be cast into
\begin{equation}
I_{v}(k_{\perp })=-\frac{\mathcal{F}(k_{\perp })}{2\pi }\left[ \left( -\frac{%
1}{\hat{\epsilon}}+\ln \frac{k_{\perp }^{2}}{\mu ^{2}}\right) C_{F}+\left(
C_{F}-\frac{N_{c}}{2}\right) \ln (1-\xi )^{2}\right] -\frac{N_{c}}{2}%
I_{3v}(k_{\perp }),
\end{equation}%
where $I_{3v}(k_{\perp })$ is finite and defined as
\begin{eqnarray}
I_{3v}(k_{\perp }) &=&2\int \frac{d^{2}q_{\perp }d^{2}k_{g1\perp }}{(2\pi
)^{2}}\mathcal{G}(k_{\perp },k_{g1\perp })\left[ \frac{q_{\perp }\cdot
\left( q_{\perp }-k_{\perp }\right) }{q_{\perp }^{2}\left( q_{\perp
}-k_{\perp }\right) ^{2}}-\frac{(q_{\perp }-\xi k_{\perp })\cdot (q_{\perp
}-k_{g1\perp })}{(q_{\perp }-\xi k_{\perp })^{2}(q_{\perp }-k_{g1\perp })^{2}%
}\right] \nonumber \\
&=&\int \frac{d^{2}k_{g1\perp }}{2\pi }\mathcal{G}(k_{\perp },k_{g1\perp
})\ln \frac{(k_{g1\perp }-\xi k_{\perp })^{2}}{k_{\perp }^{2}}.
\end{eqnarray}%
To derive the above expressions, Eq.~(\ref{intlog}) is used repeatedly. It is also useful to notice that
\begin{equation}
\int d^2k_{1\perp} e^{-ik_{1\perp}\cdot \bar{r}_{\perp}} \ln \frac{(k_{1\perp}-\xi^{\prime}
k_{\perp})^2}{k_{\perp}^2}=4\pi\left[\delta(\bar{r}_{\perp})\int \frac{d^2 r_{\perp}^{\prime}}
{r_{\perp}^{\prime2}}e^{ik_{\perp}\cdot r_{\perp}^{\prime}}-\frac{1}{\bar{r}_{\perp}^2}
e^{-i\xi^{\prime}k_{\perp}\cdot \bar{r}_{\perp}}\right], \label{momentumdelta}
\end{equation}
which can lead us to the final factorized formula.

By combining the collinear singularities from both real and virtual
diagrams, we find the coefficient of the collinear singularities becomes $%
\mathcal{P}_{qq}(\xi)$ which is defined as
\begin{equation}
\mathcal{P}_{qq}(\xi)=\left(\frac{1+\xi^2}{1-\xi}\right)_+=\frac{1+\xi^2}{%
(1-\xi)_+}+\frac{3}{2}\delta(1-\xi).
\end{equation}
Now we are ready to remove the collinear singularities by redefining the
quark distribution and the quark fragmentation function as follows
\begin{eqnarray}
q(x,\mu)&=&q^{(0)}(x)-\frac{1}{\hat{\epsilon}}\frac{\alpha_s(\mu)}{2\pi}\int^1_x%
\frac{d\xi}{\xi}C_F \mathcal{P}_{qq}(\xi)q\left(\frac{x}{\xi}\right),
\label{quarkr} \\
D_{h/q}(z,\mu)&=&D^{(0)}_{h/q}(z)-\frac{1}{\hat{\epsilon}}\frac{\alpha_s(\mu)}{%
2\pi}\int^1_z\frac{d\xi}{\xi}C_F \mathcal{P}_{qq}(\xi)D_{h/q}\left(\frac{z}{%
\xi}\right),  \label{fragr}
\end{eqnarray}
which is in agreement with the well-known DGLAP equation for the quark channel.
We will be able to recover the full DGLAP equation once we finish the
calculation on all channels. Using Eq.~(\ref{lo}) and combine it with the
NLO real and virtual contributions, it is almost trivial to show Eq.~(\ref%
{quarkr}). It is a little bit less trivial to derive Eq.~(\ref{fragr}). By
combining the relevant terms in the real and virtual contributions, we
obtain a term which reads
\begin{equation}
-\frac{1}{\hat{\epsilon}}\frac{\alpha_s(\mu)}{2\pi}\int^1_{\tau}\frac{d z}{%
z^{2}}D_{h/q}(z)\int_{\tau/z}^1 d \xi C_F\mathcal{P}_{qq}(\xi) xq(x)\frac{1}{%
\xi^2}\mathcal{F}\left(\frac{k_\perp}{\xi}\right) \ .
\end{equation}
By changing variable $z^{\prime}=z\xi$, we can rewrite the above term as
\begin{equation}
-\frac{1}{\hat{\epsilon}}\frac{\alpha_s(\mu)}{2\pi}\int^1_{\tau}\frac{d
z^{\prime}}{z^{\prime 2}} xq(x) \mathcal{F}\left(\frac{p_\perp}{z^{\prime}}%
\right) \int_{z^{\prime}}^1 \frac{d \xi}{\xi} C_F\mathcal{P}%
_{qq}(\xi)D_{h/q}\left(\frac{z^\prime}{\xi}\right),
\end{equation}
which allows us to arrive at Eq.~(\ref{fragr}) easily by combining this term
with Eq.~(\ref{lo}).

One might worry about the term which is proportional to $\frac{1}{2\pi}%
\mathcal{F}\left(k_\perp\right)\left(C_F-\frac{N_c}{2}\right)\ln (1-\xi)^2$
since it is logarithmically divergent when $\xi \to 1$. Let us show that
this singularity will also cancel between the real and virtual contributions
as follows
\begin{eqnarray}
&&\left[\int_{\tau/z}^1d\xi\frac{1+\xi^2}{(1-\xi)_+} xq(x)\ln(1-%
\xi)^2-x_pq(x_p)\int_0^1d\xi \frac{1+\xi^2}{(1-\xi)_+}\ln(1-\xi)^2\right]
\notag \\
&=&\int_{\tau/z}^1d\xi\left(\frac{(1+\xi^2)\ln(1-\xi)^2}{1-\xi}\right)_+
xq(x),  \label{log}
\end{eqnarray}
where the first term on the left hand side of the above equation comes from
the real diagrams while the second term comes from the virtual graphs. Here
we have used Eq.~(\ref{plus}) again.

\subsubsection{Finite contributions}

Now we have removed all the collinear singularities by renormalizing the
quark distribution and the quark fragmentation function. The rest of the
contribution should be finite. The last procedure is to assemble all the
finite terms into a factorized formula. For the quark channel contribution: $%
qA\rightarrow h+X$, we find that the factorization formula can be explicitly
written as
\begin{eqnarray}
\frac{d^{3}\sigma ^{p+A\rightarrow h+X}}{dyd^{2}p_{\perp }} &=&\int \frac{dz%
}{z^{2}}\frac{dx}{x}\xi xq(x,\mu )D_{h/q}(z,\mu )\int \frac{d^{2}x_{\perp
}d^{2}y_{\perp }}{\left( 2\pi \right) ^{2}}\left\{ S_Y^{(2)}(x_{\perp
},y_{\perp })\left[ \mathcal{H}_{2qq}^{(0)}+\frac{\alpha _{s}}{2\pi }%
\mathcal{H}_{2qq}^{(1)}\right] \right.   \notag \\
&&\left. +\int \frac{d^{2}b_{\perp }}{(2\pi )^{2}}S_Y^{(4)}(x_{\perp
},b_{\perp },y_{\perp })\frac{\alpha _{s}}{2\pi }\mathcal{H}%
_{4qq}^{(1)}\right\} \ ,  \label{qchannel}
\end{eqnarray}%
up to one-loop order. The leading order results have been calculated as
shown in Eq.~(\ref{lo}), from which we have
\begin{equation}
\mathcal{H}_{2qq}^{(0)}=e^{-ik_{\perp }\cdot r_{\perp }}\delta (1-\xi )\ ,
\end{equation}%
where $k_{\perp }=p_{\perp }/z$ and $r_{\perp }=x_{\perp }-y_{\perp }$. Our
objective here is to compute the hard coefficients $\mathcal{H}_{2qq}^{(1)}$
and $\mathcal{H}_{4qq}^{(1)}$. It is just straightforward to show that $%
\mathcal{H}_{2qq}^{(1)}$ reads as follows
\begin{eqnarray}
\mathcal{H}_{2qq}^{(1)} &=&C_{F}\mathcal{P}_{qq}(\xi )\ln \frac{c_{0}^{2}}{%
r_{\perp }^{2}\mu ^{2}}\left( e^{-ik_{\perp }\cdot r_{\perp }}+\frac{1}{\xi
^{2}}e^{-i\frac{k_{\perp }}{\xi }\cdot r_{\perp }}\right) -3C_{F}\delta
(1-\xi )e^{-ik_{\perp }\cdot r_{\perp }}\ln \frac{c_{0}^{2}}{r_{\perp
}^{2}k_{\perp }^{2}}\   \notag \\
&&-\left( 2C_{F}-N_{c}\right) e^{-ik_{\perp }\cdot r_{\perp }}\left[ \frac{%
1+\xi ^{2}}{\left( 1-\xi \right) _{+}}\widetilde{I}_{21}-\left( \frac{\left(
1+\xi ^{2}\right) \ln \left( 1-\xi \right) ^{2}}{1-\xi }\right) _{+}\right] ,
\label{h2qq}
\end{eqnarray}%
where the terms in the first line come from the finite logarithmic terms in $%
I_{1}(k_{\perp })$ and $I_{v}(k_{\perp })$, and $\widetilde{I}_{21}$ is
calculated from $I_{21}(k_{\perp })$ which yields
\begin{equation}
\widetilde{I}_{21}=\int \frac{d^{2}b_{\perp }}{\pi }\left\{ e^{-i\left(
1-\xi \right) k_{\perp }\cdot b_{\perp }}\left[ \frac{b_{\perp }\cdot \left(
\xi b_{\perp }-r_{\perp }\right) }{b_{\perp }^{2}\left( \xi b_{\perp
}-r_{\perp }\right) ^{2}}-\frac{1}{b_{\perp }^{2}}\right] +e^{-ik_{\perp
}\cdot b_{\perp }}\frac{1}{b_{\perp }^{2}}\right\} \ .
\end{equation}%
It is clear that the last term comes from the $\ln (1-\xi )^{2}$ terms as we
have shown in Eq.~(\ref{log}). It is also important to note that the second
line in Eq.~(\ref{h2qq}) (the $I_{2}(k_{\perp })$ type term) drops out if we
take large $N_{c}$ limit. The large $N_{c}$ will greatly simplify our
calculation in many aspects as we will show in the following sections.
Furthermore, by choosing $\mu =c_{0}/r_{\perp }$ for the factorization
scale, we can further simplify the above expressions. In the end, only the
last term in the first line of the Eq.~(\ref{h2qq}) survives. Since $%
r_{\perp }$ is of the order $1/Q_{s}$ in the saturation regime, one can
easily see that the factorization scale $\mu \simeq Q_{s}$ in terms of the
above choice.

The second hard coefficient $\mathcal{H}_{4qq}^{(1)}$ is related to the
non-linear terms such as $I_3(k_\perp)$ and $I_{3v}(k_\perp)$ which give
\begin{eqnarray}
\mathcal{H}_{4qq}^{(1)}&=&-4\pi N_{c}e^{-ik_\perp\cdot r_\perp}\left\{ e^{-i%
\frac{1-\xi}{\xi}k_\perp\cdot (x_\perp-b_{\perp})} \frac{1+\xi ^{2}}{\left(
1-\xi \right) _{+}}\frac{1}{\xi }\frac{x_\perp-b_\perp }{\left(
x_\perp-b_\perp\right) ^{2}}\cdot \frac{y_\perp-b_\perp}{\left(
y_\perp-b_\perp\right) ^{2}} \right.  \notag \\
&& \quad \left.-\delta (1-\xi )\int_{0}^{1}d\xi ^{\prime }\frac{1+\xi ^{\prime 2}}{\left(
1-\xi ^{\prime }\right) _{+}}\left[\frac{e^{-i(1-\xi^{\prime})k_\perp\cdot
(y_\perp-b_\perp)}}{(b_\perp-y_\perp)^2}-\delta^{(2)}(b_\perp-y_\perp)\int
d^2 r_\perp^{\prime}\frac{e^{ik_\perp\cdot r_\perp^{\prime}}}{%
r_\perp^{\prime 2}}\right]\right\},
\end{eqnarray}%
where the first and second term in the curly brackets are calculated from $%
I_3(k_\perp)$ and $I_{3v}(k_\perp)$, respectively.

To summarize the above results, we have demonstrated the QCD factorization for inclusive hadron production
in the quark channel of pA collisions in the saturation formalism, and we
have computed the NLO cross section in this processes. Clearly, the naive form of the $k_\perp$ factorization formula, which involves the convolution of unintegrated gluon distributions in the transverse momentum space, does not hold. Other channels can be
calculated accordingly following the same procedure.

\subsubsection{The McLerran-Venugopalan model}
In this subsection, we calculate the hard coefficients in the well-known McLerran-Venugopalan (MV) model\cite{McLerran:1993ni, Mueller:1999wm, Gelis:2001da}. In terms of the phenomenological application with additional parametrization of the saturation momentum, it is also known as Golec-Biernat-W\"{u}sthoff (GBW) model~\cite{GolecBiernat:1998js}.
In the MV and GBW model, if we neglect the impact parameter dependence for the sake of simplicity, the
dipole scattering amplitude is parametrized as
\begin{equation}
S^{(2)}_{\textrm{MV}}(x_\perp,y_\perp) =\exp \left[-\frac{(x_\perp -y_\perp)^2 Q_s^2}{4}\right] \ ,
\end{equation}
which leads to $\mathcal{F}(q_\perp)=\frac{S_\perp}{\pi Q_s^2 } \exp \left(-\frac{q_\perp^2}{Q_s^2 }\right)$,
where $S_\perp$ is the transverse area of the target hadron. In addition, if we further take the large $N_c$ limit,
the integral $d^2x_\perp d^2y_\perp d^2b_\perp$ can be performed explicitly, which leads
to the differential cross section depending on $p_\perp$ and $Q_s$,
\begin{equation}
\frac{d^{3}\sigma ^{p+A\rightarrow h+X}}{dyd^{2}p_{\perp }} =\int \frac{dz%
}{z^{2}}\frac{dx}{x}\xi xq(x,\mu )D_{h/q}(z,\mu )\left[  \bar {\mathcal{ H}}_{2qq}^{(0)}+\frac{\alpha _{s}}{2\pi }%
\bar{\mathcal{H}}_{2qq}^{(1)} +\frac{\alpha _{s}}{2\pi }\bar{\mathcal{H}}
_{4qq}^{(1)}\right] \ ,  \label{qchannelgbw}
\end{equation}
where
\begin{eqnarray}
 \bar {\mathcal{ H}}_{2qq}^{(0)}&=&\delta(1-\xi)\frac{S_\perp}{\pi Q_s^2 } \exp \left(-\frac{k_\perp^2}{Q_s^2 }\right) , \\
 \bar{\mathcal{H}}_{2qq}^{(1)}&=& \frac{N_c}{2}\mathcal{P}_{qq}(\xi ) \mathcal{F}(k_\perp) \left[ \ln \frac{Q_s^2}{\mu^2 e^{\gamma_E}} +\exp \left(\frac{k_\perp^2}{Q_s^2 }\right) L^{\left( 1,0\right) }\left( -1,-\frac{k_\perp^2}{Q_s^2}\right)\right] \nonumber \\
 &&+ \frac{1}{\xi^2}\frac{N_c}{2}\mathcal{P}_{qq}(\xi ) \mathcal{F}\left(\frac{k_\perp}{\xi}\right) \left[ \ln \frac{Q_s^2}{\mu^2 e^{\gamma_E}} +\exp \left(\frac{k_\perp^2}{\xi^2 Q_s^2 }\right) L^{\left( 1,0\right) }\left( -1,-\frac{k_\perp^2}{\xi^2 Q_s^2}\right)\right] \nonumber \\
 &&-\delta(1-\xi)\frac{3 N_c}{2} \mathcal{F}(k_\perp)  \left[ \ln \frac{Q_s^2}{k_\perp^2 e^{\gamma_E}} +\exp \left(\frac{k_\perp^2}{Q_s^2 }\right) L^{\left( 1,0\right) }\left( -1,-\frac{k_\perp^2}{Q_s^2}\right)\right] \, , \\
 \bar{\mathcal{H}}_{4qq}^{(1)}&=&-\frac{S_\perp N_c}{\pi} \frac{1+\xi^2}{(1-\xi)_+}\frac{1}{k_\perp^2} \left[1-\exp \left(-\frac{k_\perp^2}{Q_s^2 }\right)  \right] \left[1-\exp \left(-\frac{k_\perp^2}{\xi ^2 Q_s^2 }\right)  \right] \nonumber \\
 &&+ N_c\delta(1-\xi) \mathcal{F}(k_\perp) \left[\frac{3}{2} \ln \frac{Q_s^2}{k_\perp^2 e^{\gamma_E}} +\int_{0}^{1}d\xi ^{\prime }\frac{1+\xi ^{\prime 2}}{\left(
1-\xi ^{\prime }\right) _{+}}\exp \left(-\frac{\xi^{\prime 2} k_\perp^2}{Q_s^2 }\right) L^{\left( 1,0\right) }\left( -1,\frac{\xi^{\prime 2} k_\perp^2}{Q_s^2}\right)\right] \, , \label{hgbw}
\end{eqnarray}
where $k_\perp=p_\perp/z$ as above and
$L^{\left( 1,0\right) }\left( -1,-x\right) =-\left[ \gamma _{E}+\Gamma
\left( 0,-x\right) +\log \left( -x\right) \right] e^{-x}$ is the
Multivariate Laguerre Polynomial. $L^{\left( 1,0\right) }\left( -1,-x\right) $ is zero at $x=0,\infty $, and
reaches its maximum at $x$ around $2$.

An important aspect of the above results is that we can compare with the collinear
factorization results in the dilute limit. For example, the forward quark production in $pp$ collisions
is dominated by the $t$-channel $qg\to qg$ subprocess in the collinear factorization
calculation. Because of the $t$-channel dominance, we find that the differential
cross section can be written as
\begin{eqnarray}
\frac{d^3\sigma(pp\to q+x)}{d^2k_\perp dy}|_{\rm forward~limit}=
\int_{x'_{min}} \frac{dx'}{x'}xq(x) \frac{\alpha_s^2}{k_\perp^4}2x'g(x') \ ,
\end{eqnarray}
in the forward limit, where $k_\perp$ and $y$ are the transverse momentum
and rapidity of the final state quark, respectively. The above result was
obtained by taking the limit of  $-\hat t\ll \hat s\sim -\hat u$ for the Mandelstam
variables in the partonic cross section. Here, $q(x)$ and $g(x)$ are quark and gluon distributions from the incoming
two nucleons, respectively.

As a consistency check,  we can take the dilute limit which gives $k_\perp^2 \gg Q_s^2$, and
obtain the leading contribution of Eq.~(\ref{qchannelgbw}) which reads
\begin{equation}
\left.\frac{d^{3}\sigma ^{p+A\rightarrow h+X}}{dyd^{2}p_{\perp }}\right|_\textrm{dilute} =\int \frac{dz%
}{z^{2}}D_{h/q}(z,\mu ) \int \frac{d\xi}{1-\xi} xq(x,\mu )\frac{\alpha _{s}}{2\pi }\frac{2N_c S_\perp Q_s^2}{\pi k_\perp^4} . \label{qchannelgbwdilute}
\end{equation}
In arriving at the above result, we have also taken the limit $\xi \to 1$ (note that $\xi \neq 1$ for real
contributions due to subtraction) which corresponds to the limit $-\hat t\ll \hat s\sim -\hat u$. We
further notice that the quark saturation momentum~\cite{Mueller:1999wm, Mueller:2001fv}
$Q_s^2 = \frac{4\pi^2 \alpha_s}{N_c} \sqrt{R^2-b^2} \rho x' G(x')$ with $x' G(x')$
corresponding to the gluon distribution in a nucleon, $\rho$ being the nuclear density,
$R$ being the size of the target nucleus and $b$ being the impact parameter. In the
dilute regime, the gluon distribution is additive in the target nucleus which allows us to
write $x'G_A(x') = 2\sqrt{R^2-b^2} S_\perp \rho x' G(x') =A x' G(x')$ with $A$ being the
nuclear number.\footnote{Rigorously, one should write $S_\perp =\int d^2 b$ and use
the relation that $\rho \int d^2 b  2\sqrt{R^2-b^2} = \rho \frac{4\pi}{3}R^3 =A.$} At the
end of the day, by setting $\frac{d\xi}{1-\xi} =\frac{d x'}{x'}$ which recovers the integration
over the gluon longitudinal momentum fraction, we can obtain
\begin{equation}
\left.\frac{d^{3}\sigma ^{p+A\rightarrow h+X}}{dyd^{2}p_{\perp }}\right|_\textrm{dilute} =\int \frac{dz
}{z^{2}}D_{h/q}(z,\mu ) \int \frac{d x'}{x'} xq(x,\mu )\frac{2\alpha_s^2 x'G_A(x')  }{k_\perp^4} , \label{qchannelgbwdilute2}
\end{equation}
which agrees with the collinear factorization result for the quark channel.
The comparison for all other channels shall follow in the same way.
In conlusion, the factorization in Eq.~(\ref{fac}) is consistent with the collinear factorization
result in the dilute limit in the forward $pA$ collisions.

\subsection{The gluon channel $g\to g$}

The computation for the $g\to g$ channel is very similar to the calculation
we have done for the $q\to q$ channel. However, there is an additional
complication in this calculation. As we will show later in the detailed
derivation, the sextupole, namely the correlation of six fundamental Wilson
lines in a single trace, will start to appear in the cross section. The
small-$x$ evolution equation of sextupoles~\cite{Iancu:2011ns} is different
from the well-known BK equation which is derived for dipoles. This is normal
since the quadrupoles also follow a different version of small-$x$
evolution equation\cite{hep-ph/0405266, Dominguez:2011gc}. The numerical study of the evolution for sextupoles is
not yet available. Fortunately, the contribution from sextupoles is
suppressed by a factor $\frac{1}{N_c^2}$ as compared to other terms. In
addition, in principle, the four-point function $S^{(4)}(x_\perp,b_\perp,y_%
\perp)$ can not be factorized into $S^{(2)}(x_\perp,b_{\perp})S^{(2)}(b_%
\perp,y_{\perp})$ unless the large $N_c$ limit is taken. By taking the large
$N_c$ limit, not only can we simplify the calculation significantly, but
also we can show that all the relevant $S$-matrices are dipole amplitude $%
S^{(2)}$ which is universal at both leading order and NLO. From the universality
point of view, it seems that the large $N_c$ limit is essential to the factorization.
Therefore, in our
following derivation, we will take the large $N_c$ limit right away, but we
will comment on the property of the $N_c$ corrections.

\begin{figure}[tbp]
\begin{center}
\includegraphics[width=10cm]{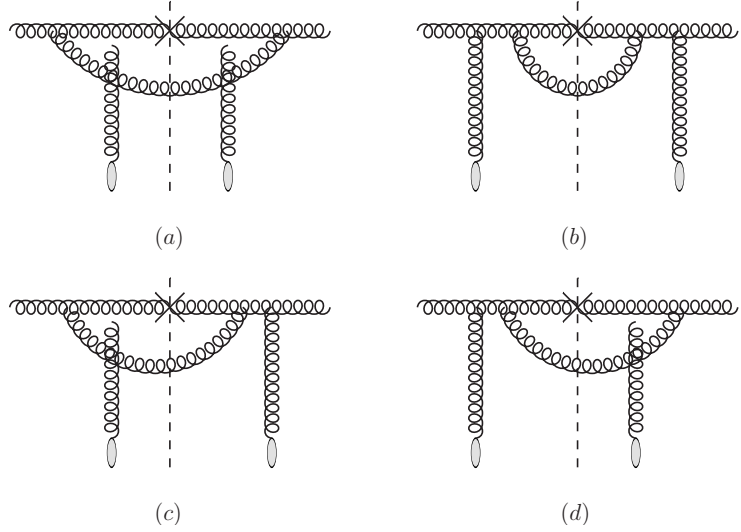}
\end{center}
\caption[*]{The real diagrams for the next-to-leading order gluon production
$gA\rightarrow g+X$.}
\label{nlogr}
\end{figure}

The real diagrams, as shown in Fig.~\ref{nlogr}, have been studied in
Ref.~\cite{Dominguez:2011wm}. Let us first analyze the $S$-matrices
associated with each graph in Fig.~\ref{nlogr}. For Fig.~\ref{nlogr}(a),
before we integrate out the phase space of the unobserved gluon, we find
that the multiple scacttering gives $\left\langle f_{ade}\left[%
W(x_\perp)W^\dagger(x^{\prime}_\perp)\right]^{db}\left[W(b_\perp)W^%
\dagger(b^{\prime}_\perp)\right]^{ec}f_{abc}\right\rangle_{Y}$, where $%
x_\perp$ and $x^{\prime}_\perp$ are the transverse coordinates of the
observed gluon in the amplitude and complex conjugate amplitude,
respectively. Here, $b_\perp$ and $b^{\prime}_\perp$ are the coordinates of the
unobserved gluon. By integrating over the phase space of the unobserved
gluon, we identify $b_\perp$ to $b^{\prime}_\perp$ which allows us to
greatly simplify the above expression and obtain $N_c\left( \left\langle%
\text{Tr}U^\dagger(x_\perp)U(x_\perp^{\prime})\text{Tr}U^\dagger(x_\perp^{%
\prime})U(x_\perp)\right\rangle_{Y}-1\right)$. The interaction
between the unobserved gluon and the nucleus target is cancelled as
expected. By taking the large $N_c$ limit, we can further drop the second term and
factorize the results into $N_cS^{(2)}(x_\perp,x_{\perp}^{\prime})S^{(2)}(x_%
\perp^{\prime},x_{\perp})$, where a factor of $\frac{1}{N_c^2}$ has been
attached as the color average~\footnote{Strictly speaking, this factor should be $%
\frac{1}{N_c^2-1}$ since the number of gluon color is $N_c^2-1$. In the
large $N_c$ limit, we just put it as $N_c^2$.}. Similarly, the Fig.~\ref%
{nlogr}(b) yields $N_cS^{(2)}(v_\perp,v^{\prime}_{\perp})S^{(2)}(v^{\prime}_%
\perp,v_{\perp})$ with $v_\perp=\xi x_\perp+(1-\xi)b_\perp$ and $%
v^\prime_\perp=\xi x^\prime_\perp+(1-\xi)b_\perp$~\footnote{The
way that we choose to define $v_\perp$ and $v^{\prime}_\perp$
here is to put the rapidity divergence at $\xi=1$ according to the
convention that the unobserved gluon's longitudinal momentum
becomes infinitely soft.}.
For Fig.~\ref{nlogr}(c), we find the scattering matrix is proportional to
\begin{eqnarray}
&&\left\langle
f_{ade}W^{db}(x_\perp)W^{ec}(b_\perp)f_{fbc}W^{af}(v^{\prime}_\perp)\right%
\rangle_{Y}  \notag \\
&=& \left\langle\text{Tr}U^\dagger(v^{\prime}_\perp)U(x_\perp)\text{Tr}U^\dagger(x_\perp)U(b_\perp)\text{Tr}
U^\dagger(b_\perp)U(v^{\prime}_\perp)
\right\rangle_{Y}  \notag \\
&&-\left\langle\text{Tr}U^\dagger(x_\perp)U(v^{\prime}_\perp)U^\dagger(b_%
\perp)U(x_\perp)U^\dagger(v^{\prime}_\perp)U(b_\perp)\right\rangle_{Y},
\label{gc}
\end{eqnarray}
where we have used Eq.~(\ref{adjoint}) and $if^{abc}T^c=[T^a,T^b]$ in the
derivation. In addition, we have assumed that the expectation value of the Wilson lines
is real, which allows us to get, for example,
\begin{eqnarray}
&&\left\langle\text{Tr}U^\dagger(v^{\prime}_\perp)U(x_\perp)\text{Tr}U^\dagger(x_\perp)U(b_\perp)\text{Tr}
U^\dagger(b_\perp)U(v^{\prime}_\perp)
\right\rangle_{Y}\nonumber \\
&=&\left\langle\text{Tr}U^\dagger(x_\perp)U(v^{\prime}_\perp)\text{Tr}%
U^\dagger(v^{\prime}_\perp)U(b_\perp)\text{Tr}U^\dagger(b_\perp)U(x_\perp)%
\right\rangle_{Y}.
\end{eqnarray}
The last term  on the right
hand side of Eq.~(\ref{gc}) is the sextupole that we discussed earlier and it is
suppressed by $\frac{1}{N_c^2}$ as compared to the first term. It is easy to
see that the first term is proportional to $N^3_c$ since it has three color
traces. Therefore, we obtain that Fig.~\ref{nlogr}(c) gives $N_cS^{(2)}(x_\perp,v^{\prime}_{%
\perp})S^{(2)}(v^{\prime}_\perp,b_{\perp})S^{(2)}(b_\perp,x_{\perp})$ in the
large $N_c$ limit. Similarly, following the same procedure, we find that
Fig.~\ref{nlogr}(d) yields $N_cS^{(2)}(v_\perp,x^{\prime}_{%
\perp})S^{(2)}(x^{\prime}_\perp,b_{\perp})S^{(2)}(b_\perp,v_{\perp})$.

Now we can follow Ref.~\cite{Dominguez:2011wm} and write down the cross
section of producing a hadron with $p_\perp$ at rapidity $y$ from a gluon as
follows
\begin{eqnarray}
\frac{d\sigma^{pA\to hX}_{\text{real}}}{d^2p_\perp dy }&=& \alpha_s
N_c\int^1_\tau \frac{dz}{z^2} D_{h/g}(z) \int^1_{\tau/z} d\xi xg(x)\int
\frac{\text{d}^{2}x_{\perp}}{(2\pi)^{2}}\frac{\text{d}^{2}x_{\perp}^{\prime }%
}{(2\pi )^{2}}\frac{\text{d}^{2}b_{\perp}}{(2\pi)^{2}}  \notag \\
&&\times e^{-ik_{\perp }\cdot(x_{\perp}-x^{\prime
}_{\perp})}\sum_{\lambda\alpha\beta}
\psi^{\lambda\ast}_{gg\alpha\beta}(u^{\prime}_{\perp})\psi^\lambda_{gg\alpha%
\beta}(u_{\perp})  \notag \\
&&\times \left[S^{(2)}(x_\perp,x_{\perp}^{\prime})S^{(2)}(x_\perp^{%
\prime},x_{\perp})+S^{(2)}(v_\perp,v^{\prime}_{\perp})S^{(2)}(v^{\prime}_%
\perp,v_{\perp})\right.  \notag \\
&&\quad\left.-S^{(2)}(x_\perp,v^{\prime}_{\perp})S^{(2)}(v^{\prime}_%
\perp,b_{\perp})S^{(2)}(b_\perp,x_{\perp})\right.  \notag \\
&&\quad\left.-S^{(2)}(v_\perp,x^{\prime}_{\perp})S^{(2)}(x^{\prime}_%
\perp,b_{\perp})S^{(2)}(b_\perp,v_{\perp})\right],  \label{partgg}
\end{eqnarray}
where the $g\to gg$ splitting kernel is found to be~\footnote{%
Here we have included the factor of $\frac{1}{p^+}$, which is in the splitting
kernel, into the cross section.}
\begin{equation}
\sum_{\lambda\alpha\beta} \psi^{\lambda\ast}_{gg\alpha\beta}(\xi,
u^{\prime}_{\perp})\psi^\lambda_{gg\alpha\beta}(\xi, u_{\perp})=4(2\pi)^2
\left[\frac{\xi}{1-\xi}+\frac{1-\xi}{\xi}+\xi(1-\xi)\right]\frac{%
u_{\perp}^{\prime}\cdot u_{\perp}}{u_{\perp}^{\prime 2} u_{\perp}^{ 2}},
\end{equation}
with $u_\perp=x_\perp -b_\perp$ and $u^{\prime}_\perp=x^{\prime}_\perp
-b_\perp$. In addition, we find that the $\xi$ dependence of the splitting
function is symmetric under the interchange $\xi \leftrightarrow (1-\xi)$,
and can be simply written as $\frac{[1-\xi(1-\xi)]^2}{\xi(1-\xi)}$. It is
clear that the real contributions contain the rapidity divergence at $\xi
\to 1$ limit.

\begin{figure}[tbp]
\begin{center}
\includegraphics[width=10cm]{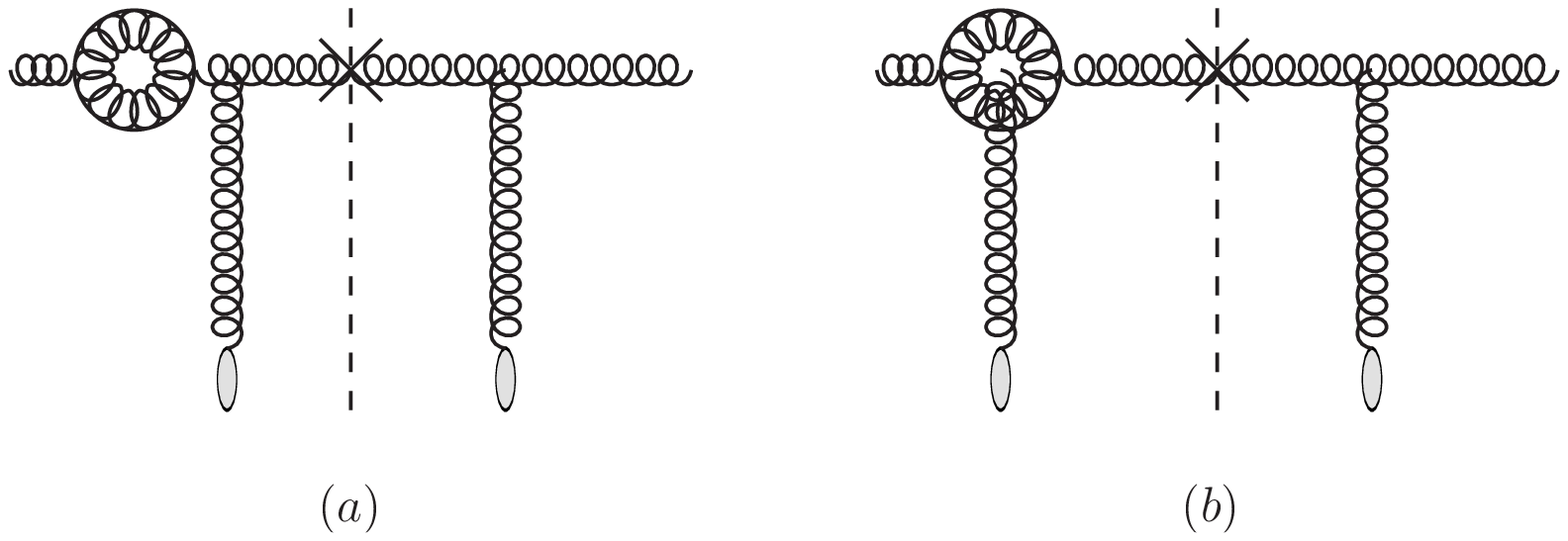}
\end{center}
\caption[*]{Typical virtual gluon loop diagrams for the next-to-leading
order gluon production $gA\rightarrow g+X$.}
\label{nlogvgluon}
\end{figure}

Similar to the quark channel, the virtual gluon diagrams as shown in
Fig.~\ref{nlogvgluon} can be calculated accordingly, and we obtain
\begin{eqnarray}
&&-\frac{2}{2}\alpha _{s}N_{c}\int_{\tau }^{1}\frac{dz}{z^{2}}%
D_{h/g}(z)x_{p}g(x_{p})\int_{0}^{1}d\xi \int \frac{d^{2}v_{\perp }}{(2\pi
)^{2}}\frac{d^{2}v_{\perp }^{\prime }}{(2\pi )^{2}}\frac{d^{2}u_{\perp }}{%
(2\pi )^{2}} \\
&&\times e^{-ik_{\perp }\cdot (v_{\perp }-v_{\perp }^{\prime
})}\sum_{\lambda \alpha \beta }\psi _{gg\alpha \beta }^{\lambda \ast
}(p^{+},\xi ,u_{\perp })\psi _{gg\alpha \beta }^{\lambda }(p^{+},\xi
,u_{\perp }) \\
&&\times \left[ S^{(2)}(v_{\perp },v_{\perp }^{\prime })S^{(2)}(v_{\perp
}^{\prime },v_{\perp })-S^{(2)}(b_{\perp },x_{\perp })S^{(2)}(x_{\perp },v_{\perp }^{\prime
})S^{(2)}(v_{\perp }^{\prime },b_{\perp })%
\right] ,  \label{v2}
\end{eqnarray}%
where the factor of $2$ comes from the fact that the mirror diagrams give
the identical contributions when the virtual loop on the right side of the
cut, while the factor of $\frac{1}{2}$ is the symmetry factor arising from two identical gluons in the closed gluon
loop. The virtual contribution contains rapidity divergence when $\xi $
approaches $0$ and $1$. This is easy to understand since the virtual gluon
loop contribution is symmetric under the interchange $\xi \leftrightarrow
(1-\xi )$. Assuming that $S^{(2)}(x_{\perp },x_{\perp }^{\prime
})=S^{(2)}(x_{\perp }^{\prime },x_{\perp })$ and use $x_{\perp }=v_{\perp
}+(1-\xi )u_{\perp }$ and $b_{\perp }=v_{\perp }-\xi u_{\perp }$, one can
easily show that the last line of is symmetric under the interchange $\xi
\leftrightarrow (1-\xi )$. Therefore, we can rewrite the splitting function $%
\left[ \frac{\xi }{1-\xi }+\frac{1-\xi }{\xi }+\xi (1-\xi )\right] $ as $2%
\left[ \frac{\xi }{1-\xi }+\frac{1}{2}\xi (1-\xi )\right] $ for the virtual
part. Now the virtual contribution only contains rapidity singularity at $%
\xi =1$.

Following the procedure that we have illustrated above for the quark
channel, we remove the rapidity divergence
terms from the real and virtual contributions by doing the following
subtractions
\begin{eqnarray}
\tilde{\mathcal{F}}(k_{\perp }) &=&\tilde{\mathcal{F}}^{(0)}(k_{\perp })-%
\frac{\alpha _{s}N_{c}}{\pi ^{2}}\int_{0}^{1}\frac{d\xi }{1-\xi }\int \frac{%
d^{2}x_{\perp }d^{2}y_{\perp }d^{2}b_{\perp }}{(2\pi )^{2}}e^{-ik_{\perp
}\cdot (x_{\perp }-y_{\perp })} \frac{(x_{\perp }-y_{\perp })^{2}}{(x_{\perp
}-b_{\perp })^{2}(y_{\perp }-b_{\perp })^{2}}  \notag \\
&&\times\left[ S^{(2)}(x_{\perp },y_{\perp })S^{(2)}(y_{\perp
},x_{\perp
})-S^{(2)}(x_\perp,y_{\perp})S^{(2)}(y_\perp,b_{\perp})S^{(2)}(b_{%
\perp },x_{\perp })\right] \ ,  \label{BKg2}
\end{eqnarray}%
where $\tilde{\mathcal{F}}^{(0)}(k_{\perp })$ is the bare dipole gluon
distribution in the adjoint representation which appears in the leading
order cross section as in Eq.~(\ref{lo}) and it is divergent. $\tilde{%
\mathcal{F}}(k_{\perp })$ is the renormalized dipole gluon in the adjoint
representation distribution and it is assumed to be finite. To arrive at
Eq.~(\ref{BKg2}), we have taken the large $N_c$ limit which allows us to
neglect the sextupole and constant term which are suppressed by $\frac{1}{%
N_c^2}$. the full subtraction should include those terms as well.

Now we are ready to show that Eq.~(\ref{BKg2}) is equivalent to the adjoint
representation of the BK equation. The non-linear small-x evolution equation
for a color dipole in some arbitrary representation $R$ can be found in
Eq.~(5.18) in Ref.~\cite{Ferreiro:2001qy}. This equation reads
\begin{eqnarray}
\frac{\partial}{\partial Y}\left\langle\text{tr}_RV^{\dagger}_{x_{%
\perp}}V_{y_{\perp}}\right\rangle_{Y}&=&-\frac{\alpha _{s}}{\pi ^{2}}\int
\frac{d^{2}z_{\perp }\,(x_{\perp }-y_{\perp })^{2}}{(x_{\perp }-z_{\perp
})^{2}(y_{\perp }-z_{\perp })^{2}}  \notag \\
&&\times \left[ C_R\left\langle\text{tr}_RV^{\dagger}_{x_{\perp}}V_{y_{%
\perp}}\right\rangle_{Y}-\left\langle\text{tr}_RV^{\dagger}_{z_{%
\perp}}t^aV_{z_{\perp}}V^{\dagger}_{x_{\perp}}t^aV_{y_{\perp}}\right%
\rangle_{Y}\right] \ ,  \label{bkr}
\end{eqnarray}
where $V$ is the Wilson line in the $R$-representation. If one
takes the fundamental representation, one can easily recover the
BK equation as shown in Eq.~(\ref{BK}). If one sets $V=W$ and uses
the adjoint representation for the color matrices
$t^a_{bc}=-if_{abc}$, one can use Eq.~(\ref{adjoint}) to convert
everything into the fundamental representation. It is
straightforward to find $C_R=N_c$ and
\begin{eqnarray}
\left\langle\text{tr}_AW^{\dagger}_{x_{\perp}}W_{y_{\perp}}\right%
\rangle_{Y}&=& \left\langle\text{Tr}U^\dagger(x_\perp)U(y_\perp)%
\text{Tr}U^\dagger(y_\perp)U(x_\perp)\right\rangle_{Y}-1  \notag \\
\left\langle\text{tr}_AW^{\dagger}_{z_{\perp}}t^aW_{z_{\perp}}W^{%
\dagger}_{x_{\perp}}t^aW_{y_{\perp}}\right\rangle_{Y}&=& \left\langle\text{Tr%
}U^\dagger(x_\perp)U(y_\perp)\text{Tr}U^\dagger(z_\perp)U(x_\perp)\text{Tr}%
U^\dagger(y_\perp)U(z_\perp)\right\rangle_{Y}  \notag \\
&&-\left\langle\text{Tr}U^\dagger(x_\perp)U(y_\perp)U^\dagger(z_\perp)U(x_%
\perp)U^\dagger(y_\perp)U(z_\perp)\right\rangle_{Y},
\end{eqnarray}
where we have also assumed all the correlation functions on the right hand
side of the above equation are real. By putting above expressions into
Eq.~(\ref{bkr}), we can obtain the adjoint representation of the BK equation
which is in complete agreement with Eq.~(\ref{BKg2}) if one also includes
the large $N_c$ corrections in Eq.~(\ref{BKg2}). This version of the BK
equation actually contains the sextupole correlation term and constant term
which coincide with Eq.~(\ref{gc}) and the discussion above. One can see
that the cancellation of the rapidity divergence is complete, even if one includes all the
large $N_c$ corrections. After the subtraction of the rapidity divergence,
the splitting functions become regulated and we can replace $\frac{1}{1-\xi}$
by $\frac{1}{(1-\xi)_+}$.

\begin{figure}[tbp]
\begin{center}
\includegraphics[width=10cm]{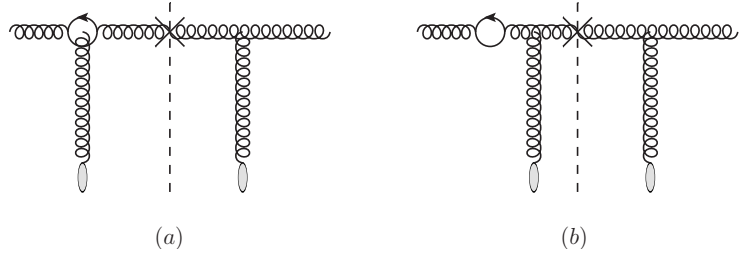}
\end{center}
\caption[*]{Typical virtual quark loop diagrams for the
next-to-leading order gluon production $gA\rightarrow g+X$.}
\label{nlogvquark}
\end{figure}

Furthermore, before we take care of the collinear singularities, we should
also compute the quark loop virtual diagrams as shown in Fig.~\ref%
{nlogvquark},
\begin{eqnarray}
&&-2\alpha _{s}N_{f}T_{R}\int_{\tau }^{1}\frac{dz}{z^{2}}%
D_{h/g}(z)x_{p}g(x_{p})\int_{0}^{1}d\xi \int \frac{d^{2}v_{\perp }}{(2\pi
)^{2}}\frac{d^{2}v_{\perp }^{\prime }}{(2\pi )^{2}}\frac{d^{2}u_{\perp }}{%
(2\pi )^{2}}  \notag \\
&&\times e^{-ik_{\perp }\cdot (v_{\perp }-v_{\perp }^{\prime
})}\sum_{\lambda \alpha \beta }\psi _{q\bar{q}\alpha \beta }^{\lambda \ast
}(p^{+},\xi ,u_{\perp })\psi _{q\bar{q}\alpha \beta }^{\lambda }(p^{+},\xi
,u_{\perp })  \notag \\
&&\times \left[ S^{(2)}(v_{\perp },v_{\perp }^{\prime })S^{(2)}(v_{\perp
}^{\prime },v_{\perp })-S^{(2)}(x_{\perp },v_{\perp }^{\prime
})S^{(2)}(v_{\perp }^{\prime },b_{\perp })\right] ,  \label{v3}
\end{eqnarray}%
where $T_{R}=\frac{1}{2}$ and the $g\rightarrow q\bar{q}$ splitting kernel
is found to be
\begin{equation}
\sum_{\lambda \alpha \beta }\psi _{q\bar{q}\alpha \beta }^{\lambda \ast
}(p^{+},\xi ,u_{\perp })\psi _{q\bar{q}\alpha \beta }^{\lambda }(p^{+},\xi
,u_{\perp })=2(2\pi )^{2}\left[ \xi ^{2}+(1-\xi )^{2}\right] \frac{1}{%
u_{\perp }^{2}}.
\end{equation}%
Normally the $g\rightarrow q\bar{q}$ channel is suppressed by one factor of $%
1/N_{c}$ as compared to other leading $N_{c}$ channels. However, the quark
loop gains a factor of $N_{f}$ since different flavors of quarks can enter
the virtual quark loop. $N_{f}$ is usually taken to be $3$ which is the same
as $N_{c}$. Therefore, we also compute this channel since this might be as
important as other contributions in terms of the numerical studies. There is
no rapidity divergence in the quark loop contribution, however, it does
contain collinear divergence.

To compute and remove the collinear divergence, we define
\begin{eqnarray}
\mathcal{K}(k_{\perp },l_{\perp },q_{\perp }) &=&\int \frac{d^{2}x_{\perp
}d^{2}b_{\perp }d^{2}y_{\perp } d^2x^\prime_\perp}{(2\pi )^{6}}e^{-ik_{\perp }\cdot (x_{\perp
}-b_{\perp })-il_{\perp }\cdot (b_{\perp }-y_{\perp })-iq_{\perp }\cdot
(y_{\perp }-x^\prime_{\perp })}  \notag \\
&&\times S^{(2)}(x_{\perp },b_{\perp })S^{(2)}(b_{\perp },y_{\perp
})S^{(2)}(y_{\perp },x^\prime_{\perp }), \label{defk}
\end{eqnarray}
where the variable $x_\perp^\prime$ is redundant and can be integrated out to
give the area of the target nucleus, if one neglects the impact factor dependence.
This allows us to transform Eq.~(\ref{partgg}) into
\begin{eqnarray}
\frac{d\sigma _{\text{real}}^{pA\rightarrow hX}}{d^{2}p_{\perp }dy} &=&\frac{%
\alpha _{s}N_{c}}{\pi ^{2}}\int_{\tau }^{1}\frac{dz}{z^{2}}%
D_{h/g}(z)\int_{\tau /z}^{1}d\xi \frac{\left[ 1-\xi \left( 1-\xi \right) %
\right] ^{2}}{\xi \left( 1-\xi \right) _{+}}xg(x)\int d^{2}q_{1\perp
}d^{2}q_{2\perp }d^{2}q_{3\perp }  \notag \\
&&\times \mathcal{K}(q_{1\perp },q_{2\perp },q_{3\perp })\left\vert \frac{%
k_{\perp }-q_{1\perp }+q_{3\perp }}{(k_{\perp }-q_{1\perp }+q_{3\perp })^{2}}%
-\frac{k_{\perp }-\xi q_{1\perp }+\xi q_{2\perp }}{(k_{\perp }-\xi q_{1\perp
}+\xi q_{2\perp })^{2}}\right\vert ^{2}.  \notag
\end{eqnarray}%
It is quite clear that among those three terms in the above equation, $\frac{%
1}{(k_{\perp }-q_{1\perp }+q_{3\perp })^{2}}$ term gives the collinear
singularity which should be absorbed by the gluon distribution, $\frac{1}{%
(k_{\perp }-\xi q_{1\perp }+\xi q_{2\perp })^{2}}$ term yields the
collinear singularity which should be associated to the
fragmentation function, while the crossing term is finite.
Similarly, the virtual gluon loop contribution can be transformed
into
\begin{eqnarray}
&&-\frac{\alpha _{s}N_{c}}{\pi ^{2}}\int_{\tau }^{1}\frac{dz}{z^{2}}%
D_{h/g}(z)x_{p}g(x_{p})\int_{0}^{1}d\xi \left[ \frac{\xi }{\left( 1-\xi
\right) _{+}}+\frac{1}{2}\xi (1-\xi )\right]   \notag \\
&&\times \int d^{2}q_{1\perp }d^{2}q_{2\perp }\mathcal{K}(q_{1\perp },q_{2\perp },q_{2\perp }-k_{\perp
})\int d^{2}l_{\perp }\left\vert \frac{%
l_{\perp }}{l_{\perp }^{2}}-\frac{l_{\perp }-\xi k_{\perp
}+q_{2\perp }-q_{1\perp }}{(l_{\perp }-\xi k_{\perp
}+q_{2\perp }-q_{1\perp })^{2}}\right\vert ^{2},
\end{eqnarray}%
and the quark loop term can be turned into%
\begin{eqnarray}
&&-\frac{\alpha _{s}N_{f}T_{R}}{2\pi ^{2}}\int_{\tau }^{1}\frac{dz}{z^{2}}%
D_{h/g}(z)x_{p}g(x_{p})\int_{0}^{1}d\xi \left[ \xi ^{2}+(1-\xi )^{2}\right]
\notag \\
&&\times \int d^{2}q_{1\perp }\mathcal{G}(q_{1\perp
},q_{1\perp }-k_{\perp })\int d^{2}l_{\perp }\left\vert \frac{l_{\perp }}{l_{\perp }^{2}}-\frac{%
l_{\perp }-\xi k_{\perp }+q_{1\perp }}{(l_{\perp }-\xi k_{\perp }+q_{1\perp })^{2}}\right\vert ^{2}.
\end{eqnarray}%
It is straightforward then to use Eq.~(\ref{loop}) to compute collinear
singularity for the virtual contributions.

By combining the collinear singularities from both real and virtual
diagrams, we find the coefficient of the collinear singularities becomes $%
\mathcal{P}_{gg}(\xi )$ which is defined as
\begin{equation}
\mathcal{P}_{gg}(\xi )=2\left[ \frac{\xi }{\left( 1-\xi \right) _{+}}+\frac{%
1-\xi }{\xi }+\xi (1-\xi )\right] +\left( \frac{11}{6}-\frac{2N_{f}T_{R}}{%
3N_{c}}\right) \delta (1-\xi ),
\end{equation}%
where the first term comes from the real diagrams, the term which is
proportional to $\frac{11}{6}\delta (1-\xi )$ comes from the virtual gluon
loop diagrams and the term which is suppressed by $1/N_{c}$ is the quark
loop contribution. Now we are ready to remove the collinear singularities by
redefining the gluon distribution and the gluon fragmentation function as
follows
\begin{eqnarray}
g(x,\mu ) &=&g^{(0)}(x)-\frac{1}{\hat{\epsilon}}\frac{\alpha _{s}(\mu )}{2\pi }%
\int_{x}^{1}\frac{d\xi }{\xi }N_{c}\mathcal{P}_{gg}(\xi )g\left( \frac{x}{%
\xi }\right) ,  \label{gluonr} \\
D_{h/g}(z,\mu ) &=&D^{(0)}{h/g}(z)-\frac{1}{\hat{\epsilon}}\frac{\alpha _{s}(\mu
)}{2\pi }\int_{z}^{1}\frac{d\xi }{\xi }N_{c}\mathcal{P}_{gg}(\xi
)D_{h/g}\left( \frac{z}{\xi }\right) ,  \label{gluonfragr}
\end{eqnarray}%
which is in agreement with the DGLAP equation for the gluon channel.

Now we are ready to assemble all the rest of the finite terms into the
hard factors. Let us take the finite terms left in the virtual contribution as
an example. Using Eq.~(\ref{loop}) to perform the $l_\perp$ integration,
the finite terms are proportional to
\begin{equation}
\int d^{2}q_{1\perp }d^{2}q_{2\perp }\mathcal{K}(q_{1\perp },q_{2\perp },q_{2\perp }-k_{\perp
}) \left[\ln\frac{k_\perp^2}{\mu^2}+\ln\frac{(l_{\perp }-\xi k_{\perp
}+q_{2\perp }-q_{1\perp })^{2}}{k_\perp^2}\right].
\end{equation}
The evaluation of the first term is trivial since it is independent of $\xi,q_{i\perp}$.
Using Eqs.~(\ref{momentumdelta}, \ref{defk}), the second term yields~\footnote{The
expression in Eq.~(\ref{v3g}) looks slightly different from the final results as shown
in Eq.~(\ref{hgg6}). Since the $S$-matrices are symmetrical among all the transverse
coordinates which are all integrated over in the end, one can exchange the definition
of variables $x_\perp \leftrightarrow y_\perp$ and reverse the orientation of all the
coordinates in Eq.~(\ref{v3g}). This allows us to show that these two expressions
are equivalent.}
\begin{eqnarray}
&&\int \frac{d^{2}x_{\perp
}d^{2}b_{\perp }d^{2}y_{\perp }}{(2\pi )^{4}}S^{(2)}(x_{\perp },b_{\perp })S^{(2)}(b_{\perp },y_{\perp
})S^{(2)}(y_{\perp },x_{\perp })e^{-ik_{\perp }\cdot (x_{\perp
}-y_{\perp })}  \nonumber \\
&&\times 4\pi\left[\delta ^{(2)}(b_{\perp
}-x_{\perp })\int d^{2}r_{\perp }^{\prime }\frac{e^{ik_{\perp }\cdot
r_{\perp }^{\prime }}}{r_{\perp }^{\prime 2}}- \frac{e^{-i\xi ^{\prime }k_{\perp }\cdot (b_{\perp
}-x_{\perp })}}{(b_{\perp }-x_{\perp })^{2}}\right]. \label{v3g}
\end{eqnarray}

Summarizing the above calculations, for the gluon channel contribution:
$gA\rightarrow h/g+X $, we find that the factorization formula can
be explicitly written as
\begin{eqnarray}
\frac{d^{3}\sigma ^{p+A\rightarrow h/g+X}}{dyd^{2}p_{\perp }} &=&\int \frac{dz%
}{z^{2}}\frac{dx}{x}\xi xg(x,\mu )D_{h/g}(z,\mu )\nonumber \\
&&\times \left\{ \int \frac{%
d^{2}x_{\perp }d^{2}y_{\perp }}{\left( 2\pi \right) ^{2}}S_{Y}^{(2)}(x_{\perp
},y_{\perp })S_{Y}^{(2)}(y_{\perp },x_{\perp })\left[ \mathcal{H}_{2gg}^{(0)}+%
\frac{\alpha _{s}}{2\pi }\mathcal{H}_{2gg}^{(1)}\right] \right.   \notag \\
&&\left. +\int \frac{d^{2}x_{\perp }d^{2}y_{\perp }d^{2}b_{\perp }}{\left(
2\pi \right) ^{4}}S_{Y}^{(2)}(x_{\perp },b_{\perp })S_{Y}^{(2)}(b_{\perp },y_{\perp
}) \frac{\alpha _{s}}{2\pi }\mathcal{H}_{2q\bar q}^{(1)}\right. \notag \\
&&\left. +\int \frac{d^{2}x_{\perp }d^{2}y_{\perp }d^{2}b_{\perp }}{\left(
2\pi \right) ^{4}}S_{Y}^{(2)}(x_{\perp },b_{\perp })S_{Y}^{(2)}(b_{\perp },y_{\perp
})S_{Y}^{(2)}(y_{\perp },x_{\perp })\frac{\alpha _{s}}{2\pi }\mathcal{H}%
_{6gg}^{(1)}\right\} \ .
\end{eqnarray}%
The leading order results have been calculated as shown in
Eq.~(\ref{lo}), from which we have
\begin{equation}
\mathcal{H}_{2gg}^{(0)}=e^{-ik_{\perp }\cdot r_{\perp }}\delta (1-\xi )\,
\end{equation}%
where $k_{\perp }=p_{\perp }/z$ and $r_{\perp }=x_{\perp }-y_{\perp }$. It
is straightforward to show that $\mathcal{H}_{2gg}^{(1)}$ and $\mathcal{%
H}_{6gg}^{(1)}$ read as follows
\begin{eqnarray}
\mathcal{H}_{2gg}^{(1)}&=&N_{c}\mathcal{P}_{gg}(\xi )\ln \frac{c_{0}^{2}}{%
r_{\perp }^{2}\mu ^{2}} \left( e^{-ik_{\perp }\cdot r_{\perp }}+\frac{1}{\xi
^{2}}e^{-i\frac{k_{\perp }}{\xi }\cdot r_{\perp }}\right) \nonumber \\
&&-\left( \frac{11}{3%
}-\frac{4N_{f}T_{R}}{3N_{c}}\right) N_{c}\delta (1-\xi )e^{-ik_{\perp }\cdot
r_{\perp }}\ln \frac{c_{0}^{2}}{r_{\perp }^{2}k_{\perp }^{2}}\ , \\
\mathcal{H}_{2q\bar q}^{(1)}&=&8\pi N_{f}T_R e^{-ik_\perp \cdot
(y_\perp -b_\perp)} \delta (1-\xi )\int_{0}^{1}d\xi ^{\prime }\left[
\xi^{\prime 2}+(1-\xi ^{\prime })^2\right]\notag \\
&& \times \left[ \frac{e^{-i\xi ^{\prime }k_{\perp }\cdot (x_{\perp
}-y_{\perp })}}{(x_{\perp }-y_{\perp })^{2}}-\delta ^{(2)}(x_{\perp
}-y_{\perp })\int d^{2}r_{\perp }^{\prime }\frac{e^{ik_{\perp }\cdot
r_{\perp }^{\prime }}}{r_{\perp }^{\prime 2}}\right],
\end{eqnarray}
and
\begin{eqnarray}
\mathcal{H}_{6gg}^{(1)} &=&-16\pi N_{c}e^{-ik_{\perp }\cdot r_{\perp
}}\left\{ e^{-i\frac{k_{\perp }}{\xi }\cdot (y-b)}\frac{\left[ 1-\xi (1-\xi )%
\right] ^{2}}{\left( 1-\xi \right) _{+}}\frac{1}{\xi ^{2}}\frac{x_{\perp
}-y_{\perp }}{\left( x_{\perp }-y_{\perp }\right) ^{2}}\cdot \frac{b_{\perp
}-y_{\perp }}{\left( b_{\perp }-y_{\perp }\right) ^{2}}\right.   \notag \\
&&\quad -\left. \delta (1-\xi )\int_{0}^{1}d\xi ^{\prime }\left[ \frac{\xi
^{\prime }}{\left( 1-\xi ^{\prime }\right) _{+}}+\frac{1}{2}\xi ^{\prime
}(1-\xi ^{\prime })\right]\left[ \frac{e^{-i\xi ^{\prime }k_{\perp }\cdot (y_{\perp
}-b_{\perp })}}{(b_{\perp }-y_{\perp })^{2}}-\delta ^{(2)}(b_{\perp
}-y_{\perp })\int d^{2}r_{\perp }^{\prime }\frac{e^{ik_{\perp }\cdot
r_{\perp }^{\prime }}}{r_{\perp }^{\prime 2}}\right] \right\} \ . \label{hgg6} 
\end{eqnarray}%
Again, by choosing $\mu =c_{0}/r_{\perp }$ for the factorization scale, we
can further simplify $\mathcal{H}_{2gg}^{(1)}$ and obtain $\mathcal{H}%
_{2gg}^{(1)}=-\left( \frac{11}{3}-\frac{4N_{f}T_{R}}{3N_{c}}\right)
N_{c}\delta (1-\xi )e^{-ik_{\perp }\cdot r_{\perp }}\ln \frac{c_{0}^{2}}{%
r_{\perp }^{2}k_{\perp }^{2}}$.

\subsection{The quark to gluon channel $q\rightarrow g$}

This channel is relatively simpler than the $q\rightarrow q$ channel
for two reasons: first, there is no virtual graphs; second, there is no
rapidity divergence in the real contributions since the lower limit of
the gluon longitudinal momentum is bounded by the hadron longitudinal momentum. It is quite
straightforward to write down the cross section for this process
by integrating out the phase space of the final state quark $%
(k_{2}^{+},k_{2\perp })$ in Eq.(\ref{partqqg}). Then, we can transform the
cross section into momentum space and take the large $N_{c}$ limit. In the
end, we obtain
\begin{eqnarray}
\frac{d\sigma _{\text{NLO}}^{pA\rightarrow h/g+X}}{d^{2}p_{\perp }dy} &=&%
\frac{\alpha _{s}N_{c}}{4\pi ^{2}}\int_{\tau }^{1}\frac{dz}{z^{2}}%
D_{h/g}(z)\int_{\tau /z}^{1}\frac{d\xi }{\xi }xq(x)\left[ 1+(1-\xi )^{2}%
\right]   \notag \\
&&\times \int d^{2}q_{1\perp }d^{2}q_{2\perp }\mathcal{G}(q_{1\perp
},q_{2\perp })\left\vert \frac{k_{\perp }-q_{1\perp }-q_{2\perp }}{(k_{\perp
}-q_{1\perp }-q_{2\perp })^{2}}-\frac{k_{\perp }-\xi q_{2\perp }}{(k_{\perp
}-\xi q_{2\perp })^{2}}\right\vert ^{2},
\end{eqnarray}%
where we have defined $\xi $ to be the longitudinal momentum fraction of the
gluon with respect to the initial quark. The production of small-$x$ gluon in $pA$
collisions has been studied quite some time ago in Ref.~\cite{Kovchegov:1998bi}.
We find complete agreement between our calculation and the partonic results in
Eqs~(56-58) of Ref.~\cite{Kovchegov:1998bi} if we remove the gluon
fragmentation function and the quark distribution,  and take the limit $\xi \to 0$
with $dy=\frac{d\xi}{\xi}$.

Following the same procedure, we remove the collinear singularities as follows%
\begin{eqnarray}
g(x,\mu ) &=&g^{(0)}(x)-\frac{1}{\hat{\epsilon}}\frac{\alpha _{s}(\mu )}{2\pi }%
\int_{x}^{1}\frac{d\xi }{\xi }C_{F}\mathcal{P}_{gq}(\xi )q\left( \frac{x}{%
\xi }\right) , \\
D_{h/q}(z,\mu ) &=&D^{(0)}_{h/q}(z)-\frac{1}{\hat{\epsilon}}\frac{\alpha _{s}(\mu
)}{2\pi }\int_{z}^{1}\frac{d\xi }{\xi }C_{F}\mathcal{P}_{gq}(\xi
)D_{h/g}\left( \frac{z}{\xi }\right) ,
\end{eqnarray}%
where we renormalize the gluon distribution and quark fragmentation function
in this off-diagonal channel. Here we have defined $\mathcal{P}_{gq}(\xi )=%
\frac{1}{\xi }\left[ 1+(1-\xi )^{2}\right] $ and substituted $\frac{N_{c}}{2}
$ by $C_{F}$ since they are equal in the large $N_{c}$ limit.

Therefore, we find the factorized cross section in this channel can be
written as
\begin{eqnarray}
\frac{d^{3}\sigma ^{p+A\rightarrow h/g+X}}{dyd^{2}p_{\perp }} &=&\int \frac{%
dz}{z^{2}}\frac{dx}{x}\xi xq(x,\mu )D_{h/g}(z,\mu )  \notag \\
&&\times \frac{\alpha _{s}}{2\pi }%
\left\{ \int \frac{d^{2}x_{\perp }d^{2}y_{\perp }}{(2\pi )^{2}}%
S_{Y}^{(2)}(x_{\perp },y_{\perp })\left[ \mathcal{H}_{2gq}^{(1,1)}+S_{Y}^{(2)}(y_{\perp },x_{\perp
})\mathcal{H}_{2gq}^{(1,2)}\right] \right.\nonumber\\
&& \left. +\int \frac{d^{2}x_{\perp }d^{2}y_{\perp
}d^{2}b_{\perp }}{(2\pi )^{4}}S_{Y}^{(4)}(x_{\perp },b_{\perp },y_{\perp })\
\mathcal{H}_{4gq}^{(1)}\right\} \ .
\end{eqnarray}%
By defining $\mathcal{W}\left( k_{1\perp },k_{2\perp }\right) =e^{-ik_{1\perp
}\cdot \left( x_{\perp }-y_{\perp }\right) -ik_{2\perp }\cdot \left(
y_{\perp }-b_{\perp }\right) }$, we find
\begin{eqnarray}
\mathcal{H}_{2gq}^{(1,1)} &=&\frac{N_{c}}{2}\frac{1}{\xi ^{2}}e^{-i\frac{%
k_{\perp }}{\xi }\cdot r_{\perp }}P_{gq}\left( \xi \right) \ln \frac{%
c_{0}^{2}}{r_{\perp }^{2}\mu ^{2}},\quad \mathcal{H}_{2gq}^{(1,2)}=\frac{N_{c}}{2}%
e^{-ik_{\perp }\cdot r_{\perp }}P_{gq}\left( \xi \right) \ln \frac{c_{0}^{2}%
}{r_{\perp }^{2}\mu ^{2}},  \notag \\
\mathcal{H}_{4gq}^{(1)} &=&-4\pi N_{c}\mathcal{W}\left( \frac{k_{\perp }}{%
\xi },k_{\perp }\right) P_{gq}\left( \xi \right) \frac{1}{\xi }\frac{%
x_{\perp }-y_{\perp }}{\left( x_{\perp }-y_{\perp }\right) ^{2}}\cdot \frac{%
b_{\perp }-y_{\perp }}{\left( b_{\perp }-y_{\perp }\right) ^{2}}.
\end{eqnarray}%
We can also choose $\mu =c_{0}/r_{\perp }$ for the factorization scale which yields $\mathcal{H}_{2gq}^{(1,1)}=\mathcal{H}_{2gq}^{(1,2)}=0$.

\subsection{The gluon channel $g\to q$}

To complete the calculation for all the channels, we should compute the $%
g\rightarrow q\bar{q},$ although it is suppressed by a factor of
$\frac{1}{N_{c}}$. For the gluon channel $g\rightarrow q\bar{q}$,
we can start from Eq. (88) of Ref.~\cite{Dominguez:2011wm}, which
allows us to obtain
\begin{eqnarray}
\frac{d\sigma _{\text{NLO}}^{pA\rightarrow h/qX}}{d^{2}k_{\perp }dy} &=&\frac{%
\alpha _{s}}{2\pi ^{2}}T_{R}\int_{\tau }^{1}\frac{dz}{z^{2}}%
D_{h/q}(z)\int_{\tau /z}^{1}d\xi xg(x)\left[ \left( 1-\xi \right) ^{2}+\xi
^{2}\right]   \notag \\
&&\times \int d^{2}q_{1\perp }d^{2}q_{2\perp }\mathcal{G}(q_{1\perp
},q_{2\perp })\left\vert \frac{k_{\perp }-\xi q_{1\perp }-\xi q_{2\perp }}{%
(k_{\perp }-\xi q_{1\perp }-\xi q_{2\perp })^{2}}-\frac{k_{\perp }-q_{2\perp }}{%
(k_{\perp }-q_{2\perp })^{2}}\right\vert ^{2}.
\end{eqnarray}%
Following the above procedure, we remove the collinear
singularities as follows%
\begin{eqnarray}
q(x,\mu ) &=&q^{(0)}(x)-\frac{1}{\hat{\epsilon}}\frac{\alpha _{s}(\mu )}{2\pi }%
\int_{x}^{1}\frac{d\xi }{\xi }T_{R}\mathcal{P}_{qg}(\xi )g\left( \frac{x}{%
\xi }\right) , \\
D_{h/g}(z,\mu ) &=&D^{(0)}_{h/g}(z)-\frac{1}{\hat{\epsilon}}\frac{\alpha _{s}(\mu
)}{2\pi }\int_{z}^{1}\frac{d\xi }{\xi }T_{R}\mathcal{P}_{qg}(\xi
)D_{h/q}\left( \frac{z}{\xi }\right) ,
\end{eqnarray}%
where we renormalize the quark distribution and gluon fragmentation function
in this off-diagonal channel. Here $\mathcal{P}_{qg}(\xi )=\left[ \left(
1-\xi \right) ^{2}+\xi ^{2}\right] $. In the end, the factorization formula
for the cross section is
\begin{eqnarray}
\frac{d^{3}\sigma ^{p+A\rightarrow h/q+X}}{dyd^{2}p_{\perp }} &=&\int \frac{%
dz}{z^{2}}\frac{dx}{x}\xi xg(x,\mu )D_{h/q}(z,\mu )\nonumber \\
&&\times \frac{\alpha _{s}}{2\pi }%
\left\{ \int \frac{d^{2}x_\perp d^{2}y_\perp }{(2\pi )^{2}}S_{Y}^{(2)}(x_{\perp },y_{\perp
})\right.\left[ \mathcal{H}_{2qg}^{(1,1)}+S_{Y}^{(2)}(x_{\perp },y_{\perp })%
\mathcal{H}_{2qg}^{(1,2)}\right]   \notag \\
&&  \left. +\int \frac{d^{2}x_\perp d^{2}y_\perp d^{2}b_\perp }{%
(2\pi )^{4}}S_{Y}^{(4)}(x_{\perp },b_{\perp },y_{\perp })\ \mathcal{H}%
_{4qg}^{(1)}\right\} \ ,
\end{eqnarray}%
with\footnote{In the dimensional regularization, the most common
convention for the gluon spin average is to use
$\frac{1}{2(1-\epsilon)}=\frac{1}{2}\left(1+\epsilon +\cdots
\right)$. The term which is proportional to $\epsilon$ can combine
with the $\frac{1}{\epsilon}$ pole terms and give a finite
contribution as seen in the second term in the square brackets in
$\mathcal{H}_{2qg}^{(1,1)} $ and $\mathcal{H}_{2qg}^{(1,2)}$.}
\begin{eqnarray}
\mathcal{H}_{2qg}^{(1,1)} &=&\frac{1}{2}e^{-ik_{\perp }\cdot r_{\perp
}}P_{qg}\left( \xi \right) \left[\ln \frac{c_{0}^{2}}{r_{\perp }^{2}\mu ^{2}} -1 \right],%
\quad \mathcal{H}_{2qg}^{(1,2)}=\frac{1}{2}\frac{1}{\xi ^{2}}e^{-i\frac{k_{\perp }%
}{\xi }\cdot r_{\perp }}P_{qg}\left( \xi \right) \left[\ln \frac{c_{0}^{2}}{
r_{\perp }^{2}\mu ^{2}}-1 \right],  \notag \\
\mathcal{H}_{4qg}^{(1)} &=&-4\pi \mathcal{W}\left( k_{\perp },\frac{k_{\perp
}}{\xi }\right) P_{qg}\left( \xi \right) \frac{1}{\xi }\frac{x_{\perp
}-y_{\perp }}{\left( x_{\perp }-y_{\perp }\right) ^{2}}\cdot \frac{b_{\perp
}-y_{\perp }}{\left( b_{\perp }-y_{\perp }\right) ^{2}}.
\end{eqnarray}%

\section{Conclusion}

In summary, we have calculated the NLO correction to inclusive
hadron production in $pA$ collisions in the small-$x$ saturation
formalism. The collinear divergences are shown to be factorized into the
splittings of the parton distribution from the incoming nucleon
and the fragmentation function for the final state hadron. As we
have shown above, the renormalization of the parton distributions
of the proton and fragmentation functions follow the well-known
DGLAP equation
\begin{equation}
\left(
\begin{array}{c}
q\left( x,\mu\right) \\
g\left( x,\mu\right)%
\end{array}%
\right) =\left(
\begin{array}{c}
q^{(0)}\left( x\right) \\
g^{(0)}\left( x\right)%
\end{array}%
\right) -\frac{1}{\hat \epsilon}\frac{\alpha \left( \mu \right)
}{2\pi }\int_{x}^{1}\frac{d\xi }{\xi }\left(
\begin{array}{cc}
C_{F}P_{qq}\left( \xi \right) &T_R P_{qg}\left( \xi \right) \\
C_{F}P_{gq}\left( \xi \right) & N_{c}P_{gg}\left( \xi \right)%
\end{array}%
\right) \left(
\begin{array}{c}
q\left(x/\xi\right) \\
g\left(x/\xi \right)%
\end{array}%
\right) ,
\end{equation}%
and
\begin{equation}
\left(
\begin{array}{c}
D_{h/q}\left( z,\mu\right) \\
D_{h/g}\left( z,\mu\right)%
\end{array}%
\right) =\left(
\begin{array}{c}
D_{h/q}^{(0)}\left( z\right) \\
D_{h/g}^{(0)}\left( z\right)%
\end{array}%
\right) -\frac{1}{\hat\epsilon}\frac{\alpha \left( \mu \right)
}{2\pi }\int_{z}^{1}\frac{d\xi }{\xi }\left(
\begin{array}{cc}
C_{F}P_{qq}\left( \xi \right) & C_{F}P_{gq}\left( \xi \right) \\
T_R P_{qg}\left( \xi \right) & N_{c}P_{gg}\left( \xi \right)%
\end{array}%
\right) \left(
\begin{array}{c}
D_{h/q}\left(z/\xi\right) \\
D_{h/g}\left(z/\xi\right)
\end{array}%
\right) ,
\end{equation}
respectively. The rapidity divergence at one-loop order is
factorized into the BK evolution in either fundamental
representation or adjoint representation for the dipole gluon
distribution of the nucleus. The hard coefficients are calculated
up to one-loop order without taking the large $N_c$ limit for the
quark $q\to q$ channel. For some technical reasons, especially
avoiding the sextupoles, as we have explained during the
derivation, we take the large $N_c$ limit for other channels. In
principle, using these hard coefficients together with the NLO
parton distributions and fragmentation functions as well as the
NLO small-$x$ evolution
equation\cite{Kovchegov:2006vj,Balitsky:2008zza} for dipole
amplitudes, one can obtain the complete NLO cross section of the
inclusive hadron production in $pA$ collisions in the large $N_c$
limit. The corrections to this NLO order cross section are either
of order $\alpha_s ^2$ or suppressed by $\frac{1}{N_c^2}$. As to
the running coupling effects~\cite{Horowitz:2010yg} in our hybrid
factorization formalism, we have no $\alpha_s$ dependence at the
leading order ($\alpha_s$ has been absorbed into the definition of
the saturation momentum),  and one power of $\alpha_s$ at the NLO,
thus we find that the one-loop approximation for the running coupling
should be sufficient.

We have shown that the differential cross section for inclusive hadron
productions in $pA$ collisions can be written in a factorization form in
the coordinate space. The factorization scale dependence in the hard
coefficients reflects the DGLAP evolutions for the quark distributions
and fragmentation functions. It is interesting to note that similar coordinate
dependence (associated with $r_\perp$) has also been found in the transverse
momentum resummation formalism derived for the Drell-Yan lepton pair production
in Ref.~\cite{CERN-TH-3923}. On the other hand, the hard coefficients in our case
do not contain double logarithms, therefore there is no need for the Sudakov
resummation for forward inclusive hadron production in $pA$ collisions.

Adding all the channels together in the large $N_c$ limit gives
\begin{eqnarray}
\frac{d^3\sigma^{ p+A\to h+X}}{dyd^2p_\perp}&=& \int \frac{dz}{%
z^2}\frac{dx}{x}
\xi \left[xq(x,\mu), xg(x,\mu)\right]\left[
\begin{array}{cc}
S_{qq}& S_{qg}\\
S_{gq} & S_{gg}%
\end{array}
\right]
\left[
\begin{array}{c}
D_{h/q}\left( z,\mu \right) \\
D_{h/g}\left(z,\mu \right)%
\end{array}%
\right] , \label{facf}
\end{eqnarray}
with factorization scale chosen as $\mu=c_0/r_\perp$ and
\begin{eqnarray}
S_{qq}&=&\int \frac{d^2x_\perp d^2y_\perp}{(2\pi)^2}S_{Y}^{(2)}(x_\perp,y_\perp)e^{-ik_\perp\cdot r_\perp}\delta(1-\xi)
\left[1-\frac{\alpha_s}{2\pi}3C_F\ln\frac{c_0^2}{r_\perp^2k_\perp^2}\right] \nonumber\\
&&+\int\frac{d^2x_\perp d^2 y_\perp d^2b_\perp}{(2\pi)^4}S_{Y}^{(4)}(x_\perp,b_\perp,y_\perp)\frac{\alpha_s}{2\pi}{\cal H}_{4qq}^{(1)}\ , \\
S_{qg}&=& \frac{\alpha_s}{2\pi}\int \frac{d^{2}x_\perp d^{2}y_\perp d^{2}b_\perp}{%
(2\pi )^{4}}S_{Y}^{(4)}(x_{\perp },b_{\perp },y_{\perp }) {\cal H}^{(1)}_{4gq}\ , \\
S_{gq}&=&\frac{\alpha_s}{2\pi}\int \frac{d^{2}x_\perp d^{2}y_\perp }{(2\pi )^{2}}S_{Y}^{(2)}(x_{\perp },y_{\perp
})\left[ \mathcal{H}_{2qg}^{(1,1)}+S_{Y}^{(2)}(x_{\perp },y_{\perp })%
\mathcal{H}_{2qg}^{(1,2)}\right]\notag\\
&&+\frac{\alpha_s}{2\pi}\int \frac{d^{2}x_\perp d^{2}y_\perp d^{2}b_\perp}{%
(2\pi )^{4}}S_{Y}^{(4)}(x_{\perp },b_{\perp },y_{\perp }) {\cal H}^{(1)}_{4qg}\ , \\
S_{gg}&=& \int \frac{%
d^{2}x_{\perp }d^{2}y_{\perp }}{\left( 2\pi \right) ^{2}}S_{Y}^{(2)}(x_{\perp
},y_{\perp })S_{Y}^{(2)}(y_{\perp },x_{\perp })e^{-ik_\perp\cdot r_\perp}\delta(1-\xi)
\left[1-\frac{\alpha_s}{2\pi}N_c\left(\frac{11}{3}-\frac{4N_fT_R}{3N_c}\right)\ln\frac{c_0^2}{r_\perp^2k_\perp^2}\right]  \notag \\
&&\left. +\int \frac{d^{2}x_{\perp }d^{2}y_{\perp }d^{2}b_{\perp }}{\left(
2\pi \right) ^{4}}S_{Y}^{(2)}(x_{\perp },b_{\perp })S_{Y}^{(2)}(b_{\perp },y_{\perp
}) \frac{\alpha _{s}}{2\pi }\mathcal{H}_{2q\bar q}^{(1)}\right. \notag \\
&&+\int \frac{d^{2}x_{\perp }d^{2}y_{\perp }d^{2}b_{\perp }}{\left(
2\pi \right) ^{4}}S_{Y}^{(2)}(x_{\perp },b_{\perp })S_{Y}^{(2)}(b_{\perp },y_{\perp
})S_{Y}^{(2)}(y_{\perp },x_{\perp })\frac{\alpha _{s}}{2\pi }\mathcal{H}%
_{6gg}^{(1)}\ ,
\end{eqnarray}
where all the hard factors are defined in previous section. Since now the
factorization scale $\mu$ depends on $r_\perp$, the parton distributions
and fragmentations function should change accordingly when we integrate
over all the coordinates. In other words, the above expression should be
understood as if the parton distributions and fragmentation functions are
written inside those coordinate integrals.

In addition, we have also demonstrated that all the hard factors can be calculated easily in
the well-known MV and GBW model and shown that our results agree with the collinear factorization results in the
dilute limit.

In the above calculations, we focus on the hadron production in the
forward $pA$ collisions, where we can safely neglect the transverse momentum
effects from the incoming parton distributions of the nucleon.
The explicit calculations at one-loop order in the above also support this
factorization, i.e., the collinear divergence associated with the incoming
parton distribution from the nucleon does not contain the transverse momentum
dependence. The situation may change if we have both small-$x$ effects
from nucleon and nucleus, such as in the mid-rapidity in $pA$ collisions at the LHC,
when the transverse momentum effects from the gluon distribution of nucleon
become important. It is in this region that a naive $k_\perp$-factorization has been
derived~\cite{Kharzeev:2003wz,Blaizot:2004wu} and
has been widely used in the literature.
It will be interesting to extend our calculations to this kinematics too.
We leave this for a future publication.

\begin{acknowledgments}
We thank E. Avsar, I. Balitsky, F. Dominguez, F. Gelis, J. Jalilian-Marian,
C. Marquet, L. McLerran, A. H. Mueller, S. Munier, J. Owens, J.W. Qiu,
A. Stasto, G. Sterman, R. Venugopalan, and W. Vogelsang
for discussions and comments.
This work was supported in
part by the U.S. Department of Energy under the contracts DE-AC02-05CH11231
and DOE OJI grant No. DE - SC0002145.
\end{acknowledgments}


\begin{thebibliography}{99}


\bibitem{Dumitru:2002qt} A.~Dumitru, J.~Jalilian-Marian,
Phys.\ Rev.\ Lett.\ \textbf{89}, 022301 (2002). 

\bibitem{Kharzeev:2003wz}
  D.~Kharzeev, Y.~V.~Kovchegov and K.~Tuchin,
  Phys.\ Rev.\  D {\bf 68}, 094013 (2003);
  D.~Kharzeev, Y.~V.~Kovchegov and K.~Tuchin,
  Phys.\ Lett.\  B {\bf 599}, 23 (2004).

\bibitem{Albacete:2003iq}
  J.~L.~Albacete, N.~Armesto, A.~Kovner, C.~A.~Salgado and U.~A.~Wiedemann,
  Phys.\ Rev.\ Lett.\  {\bf 92}, 082001 (2004).

\bibitem{Blaizot:2004wu}
  J.~P.~Blaizot, F.~Gelis and R.~Venugopalan,
  Nucl.\ Phys.\ A {\bf 743}, 13 (2004);
Nucl.\ Phys.\ \textbf{A743}, 57-91 (2004). 



\bibitem{Dumitru:2005gt} A.~Dumitru, A.~Hayashigaki, J.~Jalilian-Marian,
Nucl.\ Phys.\ \textbf{A765}, 464-482 (2006). 


\bibitem{Baier:2005dz}
  R.~Baier, Y.~Mehtar-Tani and D.~Schiff,
  Nucl.\ Phys.\  A {\bf 764}, 515 (2006).

\bibitem{JalilianMarian:2005jf}
J.~Jalilian-Marian and Y.~V.~Kovchegov,
Prog.\ Part.\ Nucl.\ Phys.\ \textbf{56}, 104 (2006).

\bibitem{Albacete:2010bs}
  J.~L.~Albacete and C.~Marquet,
  Phys.\ Lett.\  B {\bf 687}, 174 (2010).


\bibitem{Altinoluk:2011qy} T.~Altinoluk, A.~Kovner,
Phys.\ Rev.\ \textbf{D83}, 105004 (2011). 


\bibitem{Qiu:2004da}
  J.~W.~Qiu and I.~Vitev,
  Phys.\ Lett.\  B {\bf 632}, 507 (2006).

\bibitem{Guzey:2004zp}
  V.~Guzey, M.~Strikman and W.~Vogelsang,
  Phys.\ Lett.\  B {\bf 603}, 173 (2004).

\bibitem{Kopeliovich:2005ym}
  B.~Z.~Kopeliovich, J.~Nemchik, I.~K.~Potashnikova, M.~B.~Johnson and I.~Schmidt,
  Phys.\ Rev.\  C {\bf 72}, 054606 (2005).

\bibitem{Frankfurt:2007rn}
  L.~Frankfurt and M.~Strikman,
  Phys.\ Lett.\  B {\bf 645}, 412 (2007).



\bibitem{Arsene:2004ux}
  I.~Arsene {\it et al.}  [BRAHMS Collaboration],
  Phys.\ Rev.\ Lett.\  {\bf 93}, 242303 (2004).

\bibitem{Adams:2006uz}
  J.~Adams {\it et al.}  [STAR Collaboration],
  Phys.\ Rev.\ Lett.\  {\bf 97}, 152302 (2006).


\bibitem{McLerran:2011zz} L.~McLerran, J.~Dunlop, D.~Morrison and
R.~Venugopalan,
\textit{Nucl. Phys. A854 (2011) 1-256}.

\bibitem{Gribov:1984tu} L.~V.~Gribov, E.~M.~Levin and M.~G.~Ryskin,
Phys.\ Rept.\ \textbf{100}, 1 (1983). 


\bibitem{Mueller:1985wy} A.~H.~Mueller and J.~w.~Qiu,
Nucl.\ Phys.\ B \textbf{268}, 427 (1986). 

%
\bibitem{McLerran:1993ni} L.~D.~McLerran and R.~Venugopalan,
Phys.\ Rev.\ D \textbf{49}, 2233 (1994); 
Phys.\ Rev.\ D \textbf{49}, 3352 (1994). 


\bibitem{arXiv:1002.0333} F.~Gelis, E.~Iancu, J.~Jalilian-Marian and
R.~Venugopalan, 
Ann.\ Rev.\ Nucl.\ Part.\ Sci.\ \ \textbf{60}, 463 (2010).


\bibitem{Chirilli:2011km}  G.~A.~Chirilli, B.~-W.~Xiao and F.~Yuan,
Phys.\ Rev.\  Lett. {\bf 108}, 122301 (2012) [arXiv:1112.1061 [hep-ph]].



\bibitem{CU-TP-441a} A.~H.~Mueller,
Nucl.\ Phys.\ B\ \textbf{335}, 115 (1990); 
Nucl.\ Phys.\ B\ \textbf{415}, 373 (1994). 

%

\bibitem{Balitsky:1995ub} I.~Balitsky,
Nucl.\ Phys.\ \textbf{B463}, 99-160 (1996). 



\bibitem{Kovchegov:1999yj} Y.~V.~Kovchegov,
Phys.\ Rev.\ \textbf{D60}, 034008 (1999). 

%
%

\bibitem{Jalilian-Marian:1997jx+X} J.~Jalilian-Marian, A.~Kovner,
A.~Leonidov and H.~Weigert,
Nucl.\ Phys.\ B \textbf{504} (1997) 415; 
Phys.\ Rev.\ D \textbf{59} (1998) 014014:\newline
E.~Iancu, A.~Leonidov and L.~D.~McLerran,
Phys.\ Lett.\ B \textbf{510} (2001) 133; 
Nucl.\ Phys.\ A \textbf{692} (2001) 583:\newline
H.~Weigert, 
Nucl.\ Phys.\ A \textbf{703} (2002) 823. 



\bibitem{Balitsky:2010ze} I.~Balitsky, G.~A.~Chirilli,
Phys.\ Rev.\ \textbf{D83}, 031502 (2011). 

\bibitem{Beuf:2011xd}
  G.~Beuf,
  Phys.\ Rev.\ D {\bf 85}, 034039 (2012).

\bibitem{Al} A.H.~Mueller, S.~Munier, in preparation; A.H.~Mueller, private
communications.


\bibitem{Gelis:2008rw} F.~Gelis, T.~Lappi, R.~Venugopalan,
Phys.\ Rev.\ \textbf{D78}, 054019 (2008); 
Phys.\ Rev.\ \textbf{D78}, 054020 (2008). 

%

\bibitem{hep-ph/0405266} J.~Jalilian-Marian and Y.~V.~Kovchegov,
Phys.\ Rev.\ D\ \textbf{70}, 114017 (2004) [Erratum-ibid.\ D\ \textbf{71},
079901 (2005)]. 



\bibitem{Dominguez:2010xd} F.~Dominguez, B.~W.~Xiao and F.~Yuan,
Phys.\ Rev.\ Lett.\ \textbf{106}, 022301 (2011).

\bibitem{Dominguez:2011wm} F.~Dominguez, C.~Marquet, B.~W.~Xiao and F.~Yuan,
Phys.\ Rev.\ D \textbf{83}, 105005 (2011). 



\bibitem{Collins:2011zzd} J.~Collins, ``Foundations Of Perturbative QCD,''
\textit{Cambridge, UK: Univ. Pr. (2011) 624 p}

%
\bibitem{Mueller:1999wm}
  A.~H.~Mueller,
  Nucl.\ Phys.\ B {\bf 558}, 285 (1999).

\bibitem{Gelis:2001da}
  F.~Gelis and A.~Peshier,
  Nucl.\ Phys.\ A {\bf 697}, 879 (2002).

%
\bibitem{GolecBiernat:1998js}
  K.~J.~Golec-Biernat and M.~Wusthoff,
  Phys.\ Rev.\ D {\bf 59}, 014017 (1998).



\bibitem{Mueller:2001fv}
  A.~H.~Mueller,
  hep-ph/0111244.


\bibitem{Iancu:2011ns}  E.~Iancu and D.~N.~Triantafyllopoulos,
JHEP \textbf{1111}, 105 (2011); arXiv:1112.1104 [hep-ph].


%
\bibitem{Dominguez:2011gc}
  F.~Dominguez, A.~H.~Mueller, S.~Munier and B.~-W.~Xiao,
  Phys.\ Lett.\ B {\bf 705}, 106 (2011).


\bibitem{Ferreiro:2001qy}  E.~Ferreiro, E.~Iancu, A.~Leonidov and
L.~McLerran,
Nucl.\ Phys.\ A \textbf{703}, 489 (2002).

\bibitem{Kovchegov:1998bi}
  Y.~V.~Kovchegov and A.~H.~Mueller,
  Nucl.\ Phys.\ B {\bf 529}, 451 (1998).



\bibitem{Kovchegov:2006vj} Y.~V.~Kovchegov and H.~Weigert,
Nucl.\ Phys.\ A\ \textbf{784}, 188 (2007). 

\bibitem{Balitsky:2008zza} I.~Balitsky and G.~A.~Chirilli,
Phys.\ Rev.\ D \textbf{77}, 014019 (2008). 

\bibitem{Horowitz:2010yg} W.~A.~Horowitz and Y.~V.~Kovchegov,
Nucl.\ Phys.\ A\ \textbf{849}, 72 (2011). 

\bibitem{CERN-TH-3923} J.~C.~Collins, D.~E.~Soper and G.~F.~Sterman,
Nucl.\ Phys.\ B\ \textbf{250}, 199 (1985). 

\end{thebibliography}
\end{document}